\documentclass[sigconf, screen=true, nonacm]{acmart}

\usepackage{longtable}
\usepackage{capt-of}
\usepackage{balance,bm}
\usepackage{color, colortbl, xcolor}
\definecolor{LightGray}{gray}{0.97}
\usepackage{pgfplots}
\usepackage{xcolor}
\usepackage{url}

\usepackage{subcaption}
\usepackage{booktabs} 
\usepackage{graphicx}
\usepackage{textcomp}
\usepackage{color,soul}
\usepackage{bm}
\usepackage{multirow}
\usepackage{wrapfig}
\usepackage{balance}
\usepackage{graphicx}  
\usepackage{enumitem}

\usepackage{colortbl}
\usepackage{arydshln}
\usepackage [english]{babel}
\usepackage [autostyle, english = american]{csquotes}
\definecolor{linkColor}{RGB}{6,125,233}
\definecolor{green}{rgb}{0.0, 0.65, 0.31}
\definecolor{bleudefrance}{rgb}{0.19, 0.55, 0.91}
\definecolor{ceruleanblue}{rgb}{0.16, 0.32, 0.75}
\definecolor{grey}{HTML}{969696}
\definecolor{violet}{HTML}{756bb1}
\definecolor{dgrey}{HTML}{01665e}
\definecolor{lgrey}{HTML}{5ab4ac}
\definecolor{dgreen}{HTML}{005a32}
\definecolor{purple}{HTML}{ae017e}

\definecolor{editCol}{HTML}{000000}
\definecolor{maskCol}{HTML}{c51b7d}
\definecolor{lrColor}{HTML}{8856a7}
\definecolor{trColor}{HTML}{d01c8b}
\definecolor{ctColor}{HTML}{4dac26}
\definecolor{brickred}{HTML}{f03b20}
\definecolor{DarkBlue}{HTML}{00008B}
\definecolor{mscolor}{HTML}{01665e}
\definecolor{nmscolor}{HTML}{bf812d}
\definecolor{lgreen}{HTML}{ccece6}
\definecolor{dolive}{HTML}{308014}

\definecolor{editCol}{HTML}{000000}
\definecolor{maskCol}{HTML}{c51b7d}
\definecolor{lrColor}{HTML}{8856a7}
\definecolor{trColor}{HTML}{d01c8b}
\definecolor{ctColor}{HTML}{4dac26}
\definecolor{brickred}{HTML}{f03b20}
\definecolor{lgreen}{HTML}{e0f3db}
\definecolor{dpink}{HTML}{CD1076}
\definecolor{pink}{HTML}{FED2D2}
\definecolor{soothinggreen}{HTML}{4dac26}
\definecolor{darkred}{HTML}{8B0000}

\definecolor{dblue}{HTML}{215F9A}
\definecolor{violet}{HTML}{8A2BE2}
\definecolor{mscolor}{HTML}{01665e}
\definecolor{nmscolor}{HTML}{d8b365}
\definecolor{deepgrey}{HTML}{525252}
\definecolor{dslate}{HTML}{2F4F4F}
\definecolor{dolive}{HTML}{556B2F}
\definecolor{teal}{HTML}{388E8E}
\definecolor{mscolor}{HTML}{01665e}
\definecolor{nmscolor}{HTML}{d8b365}

\definecolor{aicolor}{HTML}{018571}
\definecolor{occolor}{HTML}{ff7799}

\definecolor{srcolor}{HTML}{e34a33}
\definecolor{smcolor}{HTML}{253494}
\definecolor{srsmcolor}{HTML}{7fcdbb}
\definecolor{bothcolor}{HTML}{fe9929}
\definecolor{onecolor}{HTML}{018571}
\definecolor{marroon}{HTML}{881c1c}

\usepackage{mathtools}

\usepackage{amsmath}
\usepackage{array}
\usepackage{xcolor}
\usepackage{arydshln}
\usepackage{siunitx}
\colorlet{tablerowcolor4}{gray!50} 

\definecolor{improveCol}{HTML}{7b3294}
\definecolor{worsenCol}{HTML}{008837}

\def\impbar#1{
  {\color{improveCol}\rule{#1pc}{5pt}}
  }
\def\worbar#1{
  {\color{worsenCol}\rule{#1pc}{5pt}}
  }

\definecolor{entrycol}{HTML}{CD950C}
\definecolor{exitcol}{HTML}{003057}

\def\exitbar#1{
  {\color{exitcol}\rule{#1mm}{5pt}}
  }
\def\entrybar#1{
  {\color{entrycol}\rule{#1mm}{5pt}}
  }

\newcommand*{\textlabel}[2]{%
  \edef\@currentlabel{#1}
  \phantomsection
  #1\label{#2}
}

\colorlet{tableheadcolor}{gray!25} 
\colorlet{tablerowcolor}{gray!15} 
\colorlet{tablerowcolor2}{gray!45} 
\colorlet{tablerowcolor3}{gray!25} 

\newcommand{\rowcollight}{\rowcolor{LightGray}} %
\usepackage{capt-of}
\usepackage{balance,bm}
\usepackage{color, colortbl, xcolor}

\usepackage{booktabs}  
\usepackage{url}
\usepackage{hyperref}
\hypersetup{
    colorlinks=true,
}

\usepackage{colortbl}
\usepackage{arydshln}
\usepackage [english]{babel}
\usepackage [autostyle, english = american]{csquotes}
\definecolor{linkColor}{RGB}{6,125,233}
\definecolor{green}{rgb}{0.0, 0.65, 0.31}
\definecolor{bleudefrance}{rgb}{0.19, 0.55, 0.91}
\definecolor{ceruleanblue}{rgb}{0.16, 0.32, 0.75}
\definecolor{grey}{HTML}{969696}
\definecolor{violet}{HTML}{756bb1}
\definecolor{dgrey}{HTML}{01665e}
\definecolor{lgrey}{HTML}{5ab4ac}
\definecolor{dgreen}{HTML}{005a32}
\definecolor{purple}{HTML}{ae017e}

\definecolor{editCol}{HTML}{000000}
\definecolor{maskCol}{HTML}{c51b7d}
\definecolor{lrColor}{HTML}{8856a7}
\definecolor{trColor}{HTML}{d01c8b}
\definecolor{ctColor}{HTML}{4dac26}
\definecolor{brickred}{HTML}{f03b20}
\definecolor{DarkBlue}{HTML}{00008B}
\definecolor{mscolor}{HTML}{01665e}
\definecolor{nmscolor}{HTML}{bf812d}
\definecolor{lgreen}{HTML}{ccece6}
\definecolor{dolive}{HTML}{308014}

\definecolor{editCol}{HTML}{000000}
\definecolor{maskCol}{HTML}{c51b7d}
\definecolor{lrColor}{HTML}{8856a7}
\definecolor{trColor}{HTML}{d01c8b}
\definecolor{ctColor}{HTML}{4dac26}
\definecolor{brickred}{HTML}{f03b20}
\definecolor{lgreen}{HTML}{e0f3db}
\definecolor{dpink}{HTML}{CD1076}
\definecolor{pink}{HTML}{FED2D2}
\definecolor{soothinggreen}{HTML}{4dac26}

\definecolor{dblue}{HTML}{104E8B}
\definecolor{violet}{HTML}{8A2BE2}
\definecolor{mscolor}{HTML}{01665e}
\definecolor{nmscolor}{HTML}{d8b365}
\definecolor{deepgrey}{HTML}{525252}
\definecolor{dslate}{HTML}{2F4F4F}
\definecolor{dolive}{HTML}{556B2F}
\definecolor{teal}{HTML}{388E8E}
\definecolor{mscolor}{HTML}{01665e}
\definecolor{nmscolor}{HTML}{d8b365}

\definecolor{aicolor}{HTML}{018571}
\definecolor{occolor}{HTML}{ff7799}

\definecolor{srcolor}{HTML}{e34a33}
\definecolor{smcolor}{HTML}{253494}
\definecolor{srsmcolor}{HTML}{7fcdbb}
\definecolor{bothcolor}{HTML}{fe9929}
\definecolor{onecolor}{HTML}{018571}
\definecolor{marroon}{HTML}{881c1c}

\usepackage{mathtools}

\usepackage{amsmath}
\usepackage{array}
\usepackage{xcolor}
\usepackage{arydshln}
\usepackage{siunitx}
\colorlet{tablerowcolor4}{gray!50} 

\usepackage{tcolorbox}

\colorlet{tableheadcolor}{gray!25} 
\colorlet{tablerowcolor}{gray!15} 
\colorlet{tablerowcolor2}{gray!45} 
\colorlet{tablerowcolor3}{gray!25} 

\newif{\ifhidecomments}
  \hidecommentsfalse 
\ifhidecomments
    \newcommand{\ksb}[1]{}
    \newcommand{\jackson}[1]{}
    \newcommand{\melissa}[1]{}
    \newcommand{\dongwhi}[1]{}
    \newcommand{\koustuv}[1]{}
\else
    \newcommand{\jackson}[1]{\textbf{\small\sffamily{\textcolor{DarkBlue}{[#1 -- Jackson]}}}}
    \newcommand{\melissa}[1]{\textbf{\small\sffamily{\textcolor{dolive}{[#1 -- Melissa]}}}}
    \newcommand{\dongwhi}[1]{\textbf{\small\sffamily{\textcolor{dpink}{[#1 -- Dong Whi]}}}}
    \newcommand{\ksb}[1]{\textbf{\small\sffamily{\textcolor{marroon}{[#1 -- Karthik]}}}}
    \newcommand{\koustuv}[1]{\textbf{\small\sffamily{\textcolor{violet}{[#1 -- Koustuv]}}}}
  \fi
\newcommand{\edit}[1]{{\textcolor{editCol}{#1}}}

\newcommand{\ws}{\texttt{WellScreen}}
\newcommand{\ea}{E--A gap}

\newcommand{\crea}{\textit{Creativity}}
\newcommand{\ent}{\textit{Entertainment}}
\newcommand{\pro}{\textit{Productivity}}
\newcommand{\shop}{\textit{Shopping}}
\newcommand{\soc}{\textit{Social}}

\renewcommand{\textrightarrow}{$\rightarrow$}

\newcommand{\n}[1]{$\mathtt{#1}$}

\colorlet{tableheadcolor}{gray!25} 
\colorlet{tablerowcolor}{gray!5} 

\definecolor{neutralCol}{HTML}{dd1c77}
\definecolor{neutralGreen}{HTML}{31a354}
\definecolor{NewBlue}{HTML}{1879ba}
\definecolor{bleudefrance}{rgb}{0.19, 0.55, 0.91}  
\definecolor{AfTrColor}{HTML}{0868ac}  
\definecolor{BfTrColor}{HTML}{a8ddb5}  

\definecolor{AfCtColor}{HTML}{b10026}  
\definecolor{BfCtColor}{HTML}{fd8d3c}

\graphicspath{ {figures/} }

\newcommand{\para}[1]{\vspace{0.3em}\noindent\textbf{#1}~}

\AtBeginDocument{%
  \providecommand\BibTeX{{%
    \normalfont B\kern-0.5em{\scshape i\kern-0.25em b}\kern-0.8em\TeX}}}

\begin{document}

\title[Designing Toward Digital Self-Awareness and Digital Wellbeing]{``In my defense, only three hours on Instagram'':\\ Designing Toward Digital Self-Awareness and Wellbeing}


\author{Karthik S. Bhat}
\orcid{0000-0003-0544-6303}
\affiliation{%
  \institution{Drexel University}
 \city{Philadelphia}
 \state{PA}
 \country{USA}}
 \email{ksbhat@drexel.edu}

\author{Jiayue Melissa Shi}
\orcid{0009-0007-0624-2421}
\authornote{Both authors contributed equally.}
\affiliation{%
  \institution{University of Illinois Urbana-Champaign}
 \city{Urbana}
 \state{IL}
 \country{USA}}
 \email{mshi24@illinois.edu}

\author{Wenxuan Song}
\orcid{0009-0006-0718-6882}
\authornotemark[1]
\affiliation{%
  \institution{University of Illinois Urbana-Champaign}
 \city{Urbana}
 \state{IL}
 \country{USA}}
 \email{js129@illinois.edu}

\author{Dong Whi Yoo}
\orcid{0000-0003-2738-1096}
\affiliation{%
 \institution{Indiana University Indianapolis}
 \city{Indianapolis}
 \state{IN}
 \country{USA}}
 \email{dy22@iu.edu}
 
\author{Koustuv Saha}
\orcid{0000-0002-8872-2934}
\affiliation{%
  \institution{University of Illinois Urbana-Champaign}
  \city{Urbana}
  \state{IL}
  \country{USA}}
\email{ksaha2@illinois.edu}

\renewcommand{\shortauthors}{Bhat et al.}



\begin{abstract}

Screen use pervades daily life, shaping work, leisure, and social connections while raising concerns for digital wellbeing. Yet, reducing screen time alone risks oversimplifying technology’s role and neglecting its potential for meaningful engagement. We posit self-awareness---reflecting on one’s digital behavior---as a critical pathway to digital wellbeing. We developed \ws{}, a lightweight probe that scaffolds daily reflection by asking people to estimate and report smartphone use. In a two-week deployment \edit{with college students} (\n{N}=25) \edit{focused on generating formative insights}, we examined how discrepancies between estimated and actual usage shaped digital awareness and wellbeing. Participants often underestimated productivity and social media while overestimating entertainment app use. They showed a 10\% improvement in positive affect, rating \ws{} as moderately useful. Interviews revealed that structured reflection supported recognition of patterns, adjustment of expectations, and more intentional engagement with technology. Our findings highlight the promise of lightweight reflective interventions for supporting self-awareness and intentional digital engagement, offering implications for designing digital wellbeing tools.

\end{abstract}

\ccsdesc[300]{Human-centered computing~Empirical studies in collaborative and social computing}
\ccsdesc[300]{Applied computing~Psychology}

\keywords{wellbeing, digital wellbeing, screen time, expectation--reality gap}

\maketitle


\section{Introduction}\label{section:intro}






The rapid growth of digital technology has fundamentally transformed how people work, communicate, and navigate daily life. 
Smartphones, laptops, wearables, and online platforms are no longer optional conveniences but essential tools for education, employment, and social connection---and living an entirely \textit{analog} life is virtually impossible. 
The ubiquity of digital tools has also sparked growing concerns about their impact on our productivity and wellbeing. 
Research has linked heavy or unregulated technology use to lower psychological wellbeing, stress, and reduced satisfaction with life~\cite{przybylski2017large,twenge2018associations,haidt2024anxious}, while broader critiques point to recurring ``technology panics'' in public discourse~\cite{orben2020sisyphean}.
These concerns have fueled the emergence of the field of \textit{digital wellbeing} (DW), which broadly considers how individuals can live healthy, balanced lives in increasingly digital environments~\cite{burr2020ethics,vanden2021digital,monge2019race}. 
Within human–computer interaction (HCI), a wide range of approaches have been explored to promote DW, from interventions that track and visualize screen time to systems that enforce self-control through timers, goal-setting, or punishment–reward mechanisms~\cite{ko2015nugu,zhou2021time,lyngs2019self,nguyen2021managing}. 
While such approaches are beneficial, they also predominantly tend to frame DW as a matter of restriction---using less technology, spending fewer hours online, or controlling impulses.

Reducing technology use through restricting screen time alone, however, is neither a sufficient nor always a practical measure for supporting wellbeing.
While~\citeauthor{haidt2024anxious} argues that digital technologies, particularly social media, cause declining youth mental health, critics caution that such claims may overstate the evidence and risk oversimplifying complex social and developmental contexts~\cite{odgers2024great,odgers2024great,taylor2024adolescents}.
Scholars such as~\citeauthor{odgers2024great} have argued that associations between screen time and adolescent wellbeing are small, inconsistent, and heavily dependent on methods and measures~\cite{odgers2024great,taylor2024adolescents}. 
In practice, the boundaries between ``productive'' and ``leisure'' technology use are blurry, as the same device may be used within minutes for work, communication, entertainment, and creative expression. 
As such, focusing solely on \textit{restriction} can lead to overlooking the diverse ways digital technologies can meaningfully contribute to an individual's wellbeing.
Recent research shows that technology use is not inherently harmful; its value is derived from how it is used and whether it supports meaningful engagement, personal goals, and mental health~\cite{saha2025mental}.
This shift highlights a critical need to move beyond oversimplified notions of screen time, and 
for designing toward mindful and meaningful technology use, promoting intentional, value-driven engagement~\cite{lukoff2018makes,thatcher2018mindfulness}.

We argue that \textit{self-awareness}---defined as the capacity to attend to and reflect upon one's own behaviors, emotions, and experiences~\cite{duval1972theory}---could be a critical but underexplored pathway toward DW. In the context of digital interactions, self-awareness extends to being mindful of one's technology use, recognizing its presence in---and effects on---everyday life. 
We posit that designing towards greater self-awareness could improve an individual's DW. 
A key dimension in facilitating such self-awareness is the \textit{estimated--actual gap} (\ea{}), denoting the gulf between a user's perceived or estimated use of digital technologies and their actual use. \ea{} has been linked to feelings of lost control and lower satisfaction~\cite{boase2013measuring,ernala2020well,ellis2019smartphone,parry2021systematic}. 
From this perspective, self-awareness and reconciliation of one's \ea{} could aid in fostering a greater sense of control and satisfaction in technology use, enhancing one's DW. 
We further emphasize designing technologies that \textit{foster reflection and mindful recognition of digital habits} as a means toward DW. Such awareness could help individuals make more intentional choices about their technology engagement. By cultivating greater awareness of such discrepancies and personal tendencies, individuals may be better positioned to develop healthier relationships with technology---not merely by using it less, but by using it more mindfully and meaningfully.

Despite the promise of this perspective, little is known about how self-awareness of digital behavior develops in everyday contexts, or how lightweight interventions might scaffold it. This paper aims to advance understanding of this issue by examining how regular self-reflection on users' estimated--actual gap is associated with their technology use behaviors and their digital wellbeing. Specifically, our work is guided by the following research questions (RQs):

\begin{enumerate}
    \item[\textbf{RQ1:}] How do users' estimations of digital use compare with their actual use, and what factors explain the disparities?
    \item[\textbf{RQ2:}] How does a self-reflection tool influence users' self-reported wellbeing and understanding of digital wellbeing?
    \item[\textbf{RQ3:}] How do users perceive the usefulness of a self-reflection tool, and how does it support digital self-awareness?
    
\end{enumerate}

\edit{Being a preliminary study into the usefulness of self-reflection for digital wellbeing,} we scoped our study to a population of college students. 
As heavy technology users---particularly smartphones---balancing academic, personal, and social demands, college students are both highly dependent on digital tools and vulnerable to their challenges. 
Therefore, understanding how college students perceive and reflect on smartphone use can yield insights into designing tools that emphasize awareness and mindful engagement.

We built a technology probe called \ws{}, a digital assistant that scaffolds daily self-reflection on smartphone use by prompting participants to estimate their daily usage at the start of the day, update their estimation at the end of the day, report their actual usage from their device's automatically tracked metrics, and compare these with reflective self-assessments~\cite{hutchinson2003technology}. 
Through its structured, lightweight interactions, \ws{} aimed to make these gaps visible and support mindful engagement with technology.

We recruited 25 participants for a two-week deployment study with \ws{} along with surveys and interviews at entry and exit.
Participants often misestimated their use---typically underestimating productivity and social media while overestimating entertainment. Regression analyses showed that lower \ea{}s were associated with greater satisfaction and self-control, but also higher stress and difficulty of goal adherence.
Participants a significant 10\% increase in positive affect, and they rated \ws{} as moderately useful (median SUS=80, IAM=14).
Interviews explained and complemented the above results, highlighting how structured self-reflection through \ws{} enabled participants to better understand their smartphone usage patterns and develop more nuanced understandings of their digital wellbeing. Our findings also surfaced smartphone usage behaviors that might have negatively influenced our participants' mental and emotional wellbeing, which they then critically reflected on in their daily logs by associating positive or negative justifications to their \ea{}.

Together, our study underscores the need for HCI to design digital wellbeing tools that move beyond restrictive screen time controls and instead support reflective practices that integrate emotional, social, and contextual dimensions of technology use, enabling users to develop more mindful and personally meaningful relationships with technology. This paper contributes: 1) empirical evidence demonstrating the promise of lightweight interventions in scaffolding self-reflections for digital wellbeing; 2) insights into how users develop a more holistic understanding of wellbeing---including digital, emotional, and social dimensions---through reflective engagement; and 3) design recommendations for technologies that can enable digital self-awareness through estimation--reflection workflows and data interrogation.

\section{Background and Related Work}\label{section:rw}

\subsection{Digital Wellbeing: Concept and Importance}


Digital Wellbeing is an emerging area of research broadly investigating the interplay between digital technologies and individuals' physical, mental, and emotional wellbeing in an information society~\cite{Burr_Analysis_2018}. This topic has received growing attention from a variety of disciplines including among scholars in behavioral and mental health~\cite{Andreassen_relationship_2017}, psychology~\cite{Turkle_Alone_2011}, and HCI~\cite{roffarello2023achieving}. We describe how we situate our research within this body of prior work below.


\subsubsection{Digital Wellbeing in HCI}

Human-Computer Interaction (HCI) research on digital wellbeing has grown significantly in response to concerns about the effects of pervasive technology use on users' mental health, focus, and autonomy~\cite{cecchinato2019designing,monge2021coping}. Initial work focused on screen time and overuse, leading to the development of tools that monitor and limit usage~\cite{eichner2020planting}. \citet{eichner2020planting} provided an academic taxonomy of digital wellbeing app features, characterizing them as tools to help users track, analyze, and limit their digital consumption. ~\citeauthor{Roffarello_Race_2019} analyzed 42 digital wellbeing apps and found that while users appreciate these tools, they primarily rely on self-monitoring approaches that fail to promote lasting behavioral change. In particular, users adhered to only 2\% of the apps' timer-based intervention~\cite{Roffarello_Race_2019}. In the recent past, \citeauthor{Cecchinato_Designing_2019} organized a workshop at CHI to establish a research agenda for digital wellbeing, arguing that the field needs to move beyond medicalized framings of technology ``addiction'' toward understanding how to support meaningful and intentional technology interactions~\cite{Cecchinato_Designing_2019}. Indeed, subsequent research has highlighted that digital wellbeing is more than managing screen time--it includes emotional, cognitive, and behavioral dimensions of users' relationships with technology~\cite{roffarello2023achieving}. Researchers have explored designed strategies such as persuasive interfaces, friction, and value-sensitive design to promote intentional and reflective use, noting that applications with micro-incentives can contribute to unhealthy attachments~\cite{Churchill_2019}. 

Importantly, one-size-fits-all solutions often misalign with users' evolving goals and contexts, as they may fail to align with a user's instrumental or habitual attachment to their devices~\cite{lukoff2018makes}. Recent work emphasized user-centered frameworks that promote agency, flexibility, and meaningful engagement, expanding the scope of digital wellbeing beyond usage reduction toward fostering sustainable and personally meaningful technology experiences~\cite{Grosse-Hering_Slow_2013, Zhu_Designing_2017}.




\subsubsection{Digital Wellbeing for College Students}

College students---a demographic consisting of young adults---are among the most prolific users of digital technologies, particularly smartphones and social media platforms~\cite{twenge2018associations, przybylski2019digital}. 
In the U.S., 50.4\% of teenagers reported 4 or more hours of non-school screen time per weekday from mid-2021 to late-2023~\cite{zablotsky2024daily}. Longitudinal estimates indicate an upward trend: American teens' average screen time rose from 6.5 hours in 2015 to above 8.5 hours by 2021~\cite{Duarte2025ScreenTimeTeens}.
In particular, since the COVID-19 pandemic, the use of remote and digital technologies has proliferated, further blurring the boundaries between work and non-work~\cite{saha2025mental2,farivar2024constant,chinyamurindi2022intended}.
Together, these figures highlight that college students often engage with screens more intensely over recent years. 

While higher engagement with digital technology provides opportunities for learning, self-expression, and social connection, it can also impact wellbeing. Excessive or unbalanced use has been linked to disrupted sleep, decreased attention, and negative affect in college student populations~\cite{twenge2018associations, przybylski2019digital}. Importantly, college students' developmental stage---marked by identity formation and heightened sensitivity to peer feedback---may intensify the emotional and cognitive effects of online interactions~\cite{przybylski2019digital}. Moreover, the estimated--actual gap in device use appears particularly pronounced among students, with self-estimates frequently diverging from behavioral usage data~\cite{roffarello2023achieving}. 

These studies highlight that college students engage with digital technologies more intensely than other populations, and that this engagement can have measurable cognitive, emotional, and behavioral effects. Prior research has documented screen time trends, associated risks, and expectation-reality gaps in self-reported technology use, but less is known about how college students' actual usage patterns interact with their self-perceptions in everyday contexts~\cite{judice2023discrepancies, goetzen2023likes}. Our study builds on this body of work by examining these dynamics in real-world settings, providing insights into the challenges and opportunities for supporting digital wellbeing. 



\subsection{Self-Awareness and Self-Reflective Tools}




Self-awareness is a concept that encapsulates how people deliberately attend to their own thoughts and feelings and how they engage with those around them~\cite{sutton2016measuring}. In digital wellbeing, self-awareness has been leveraged to examine one's own technology-use patterns and assessing their impact on personal behavior change~\cite{almourad2021digital}. 

Prior work in HCI has explored various technologies to support this self-awareness---through approaches like self-tracking (\textit{e.g.,} \cite{lallemand_trinity_2024, murnane_personal_2018,lee2025peerspective}) and self-reflection (\textit{e.g.,} \cite{arnera_digital_2024, pammer_reflection_2021, Kim_TimeAware_2016})---among users . \citeauthor{roffarello2023achieving} conducted a systematic review and meta-analysis and found that digital self-control tools such as self-timers achieve \textit{``a small to medium effect on reducing the time spent by users on distractive technological sources,''} while calling for longer studies and more user-centered design guidance~\cite{roffarello2023achieving}. \citeauthor{zhang2022monitoring} discovered that simple usage of dashboards and pop-up nudges makes people aware of their screen time but does not help them feel more in control~\cite{zhang2022monitoring}. In response to these limitations, other systems have focused on enabling more deliberate user engagement. For example, \citeauthor{terzimehic2023mixed} introduces Rabbit Hole Tracker, a smartphone app that logs each usage session and, at every unlock and lock event, prompts users first for their intended purpose and later for whether they completed or exceeded that intention—information that reveals when actual behavior departs from the original plan~\cite{terzimehic2023mixed}. 
Prior work has also examined the effectiveness of computer-supported time protection to minimize distractions and enhance focused work~\cite{dasswain2023focused,saha2023focus,grover2020design,kimani2019conversational,tseng2019overcoming}. 
Additionally, \citeauthor{purohit2020designing} introduced NUDGE, a digital nudge tool that prompts users to reflect on their social media use in real time, helping users become more reflective and possibly reduce usage~\cite{purohit2020designing}.

Beyond self-tracking and reflection tools, researchers have also investigated direct interventions aimed at reducing excessive screen use through mechanisms such as app blockers, lockout features, and friction-based interface design that disrupts habitual interaction~\cite{almoallim2022toward, zheng2024soap, renner2016effects}. For instance, \citeauthor{hiniker2016mytime} designed the MyTime app, which enabled users to set personal rules for phone usage, resulting in more mindful engagement and reduced daily screen time~\cite{hiniker2016mytime}. Similarly, \citeauthor{ko2016lock} developed Lock n' LoL, a smartphone application that lets group members lock their phones at the same time, creating synchronous awareness that everyone is limiting use and helping the whole group stay focused on the shared activity~\cite{ko2016lock}. More recently, \citeauthor{haliburton2024longitudinal} conducted a large-scale study of the One Sec app, which inserts short delays before opening selected applications, creating friction that interrupts automatic behaviors and encourages users to reconsider their actions~\cite{haliburton2024longitudinal}. While these interventions could be effective in limiting immediate screen exposure, many may fall short in fostering deeper self-awareness or connecting use with personal values and goals~\cite{almoallim2022toward}. This challenge has led to a growing shift toward systems that cultivate internal motivation and reflective thinking rather than relying solely on restrictive controls, and we aim to contribute to this body of work in this paper.

Building on the established role of technology probes in generating insights from real-world user data, our work with \ws{} contributes to this body of literature by introducing a specific design that leverages the gap between estimation and actual use~\cite{hutchinson2003technology}. By presenting these data points side-by-side, our probe provides a novel mechanism for scaffolding self-reflection and cultivating digital self-awareness, which is important and underexplored, as a pathway to digital wellbeing. This approach moves beyond simple monitoring to empower users to recognize their own behavioral patterns and consciously adjust their habits, offering clear design implications for future reflective interventions.

\subsection{Human-Centered Design of Wellbeing Tools}




Human-centered design has become central to developing wellbeing technologies that are not only effective but also align with users' needs~\cite{shen2022human}. Across the field, participatory and co-design methods have been used to ensure that wellbeing tools are grounded in the real-world challenges of target populations~\cite{mcgovern2025use, kilfoy2024umbrella, malloy2022co}. \citeauthor{poot2023use} used a participatory design process with children and their caregivers to create Hospital Hero, an eHealth app that seeks to ease children's pre-procedure anxiety through story-driven interaction flows tailored to their emotional needs~\cite{poot2023use}. Besides that,~\citeauthor{van2021designing} used a five-step participatory-design process with young adults to co-create an eHealth wellbeing programme whose content and delivery mirror their daily contexts, motivations, and routines \cite{van2021designing}. Complementing these participatory efforts, personal informatics research has highlighted how self-tracking systems succeed only when they support reflection \cite{li2010stage} and fit into the lived routines of users \cite{rapp2016personal}. Together, these studies demonstrate that aligning design with the everyday experiences of users fosters deeper engagement and increases the likelihood that wellbeing tools will be meaningfully adopted.

In addition to participatory and co-design practices, researchers have emphasized the need to incorporate user perceptions, trust, and autonomy into the development of wellbeing technologies~\cite{chamorro2021self,moilanen2023supporting}.~\citeauthor{choe2014understanding} found that many self-trackers stopped using their tools when the manual logging left them ``really fatigued'' and when the systems offered no ``proper tools to interpret the significance'' of the data they had collected, making meaningful insight-building too burdensome~\cite{choe2014understanding}. Furthermore,~\citeauthor{rahman2025assessing} created and tested persuasive system prototypes that included transparency, adjustable autonomy, and informed consent features~\cite{rahman2025assessing}. 
Together, these findings highlight that beyond involving users in design, it is essential to build systems that give users a sense of control, encourage meaningful interpretation, and foster long-term trust to sustain engagement with wellbeing technologies. 
We build on these insights to design a lightweight intervention, \ws{}, aimed at examining how self-reflection can support self-awareness of smartphone use.

\section{Study Design and Methods}


The goal of our research was to examine how the cognizance of, and regular self-reflection on, college students' estimated--actual gap in smartphone usage habits could advance our understanding of digital self-awareness and digital wellbeing (DW). We describe our approach and methods here.
Our study received Institutional Review Board (IRB) approval from our universities and took place between April and August 2025.


\subsection{Developing a technology probe: \ws{}} 

For the purposes of our study, we built a technology probe called \ws{}---a digital assistant designed to facilitate participants' daily self-reflections on their smartphone usage in a way that is both structured and user-friendly~\cite{hutchinson2003technology}. 
The primary purpose of any technology probe is to collect user data in real-world settings to better understand how people use technologies and how future technologies can better support them~\cite{hutchinson2003technology}. 
In the context of DW, such probes are particularly valuable because they not only capture patterns of device use but also prompt participants to engage in self-reflection, surfacing the gap between perceived and actual behaviors~\cite{wang2021salientrack}. 
This dual role allows a technology probe like \ws{} to generate insights into both user practices and design opportunities for future interventions. 

\begin{figure*}[t!]
 \begin{subfigure}[b]{0.66\columnwidth}
    \centering
    \includegraphics[width=\columnwidth]{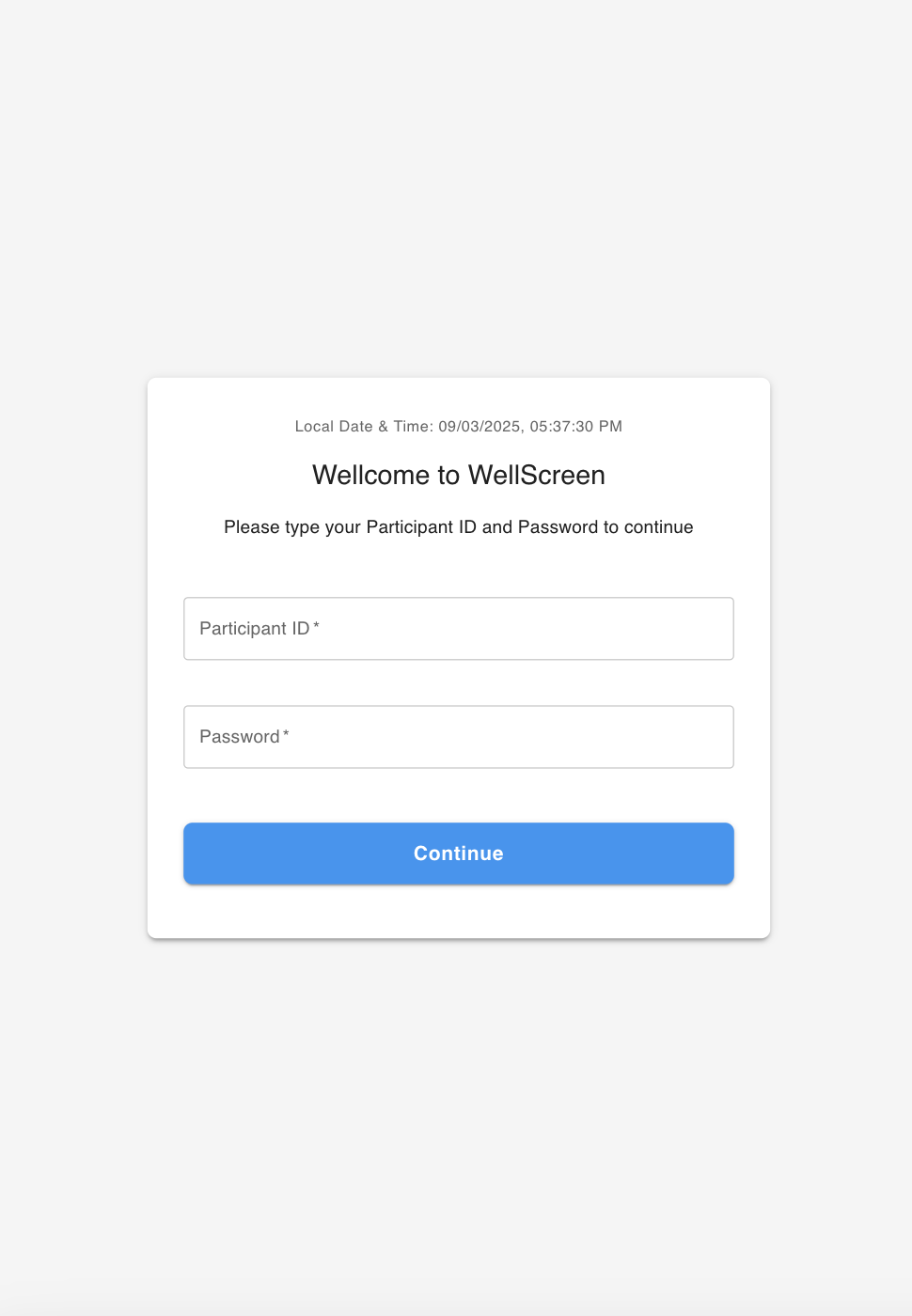}
    \caption{Login page}
    \end{subfigure}\hfill
\begin{subfigure}[b]{0.66\columnwidth}
    \centering
    \includegraphics[width=\columnwidth]{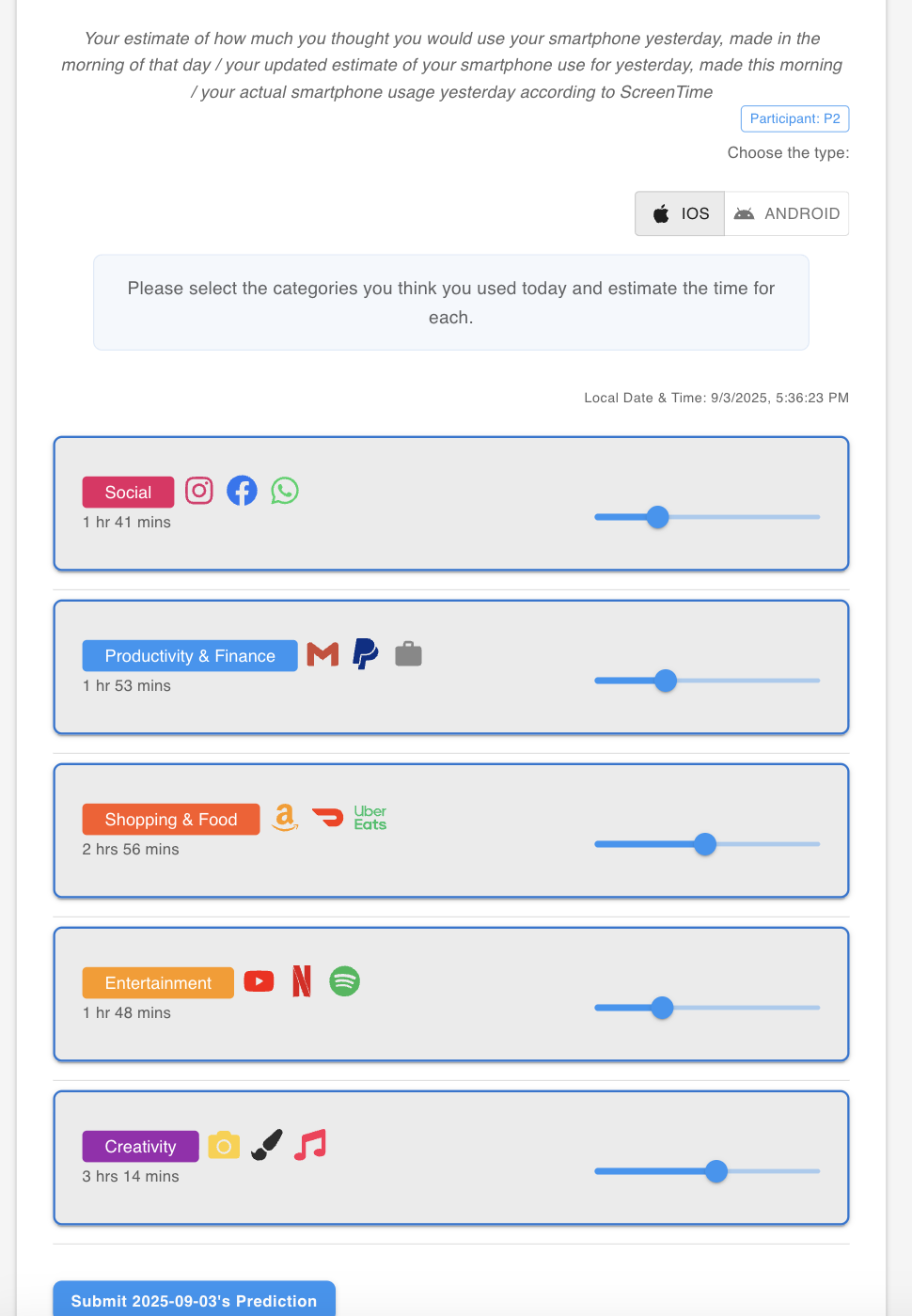}
    \caption{Start-of-the-day Estimation}
    \end{subfigure}\hfill
 \begin{subfigure}[b]{0.66\columnwidth}
    \centering
    \includegraphics[width=\columnwidth]{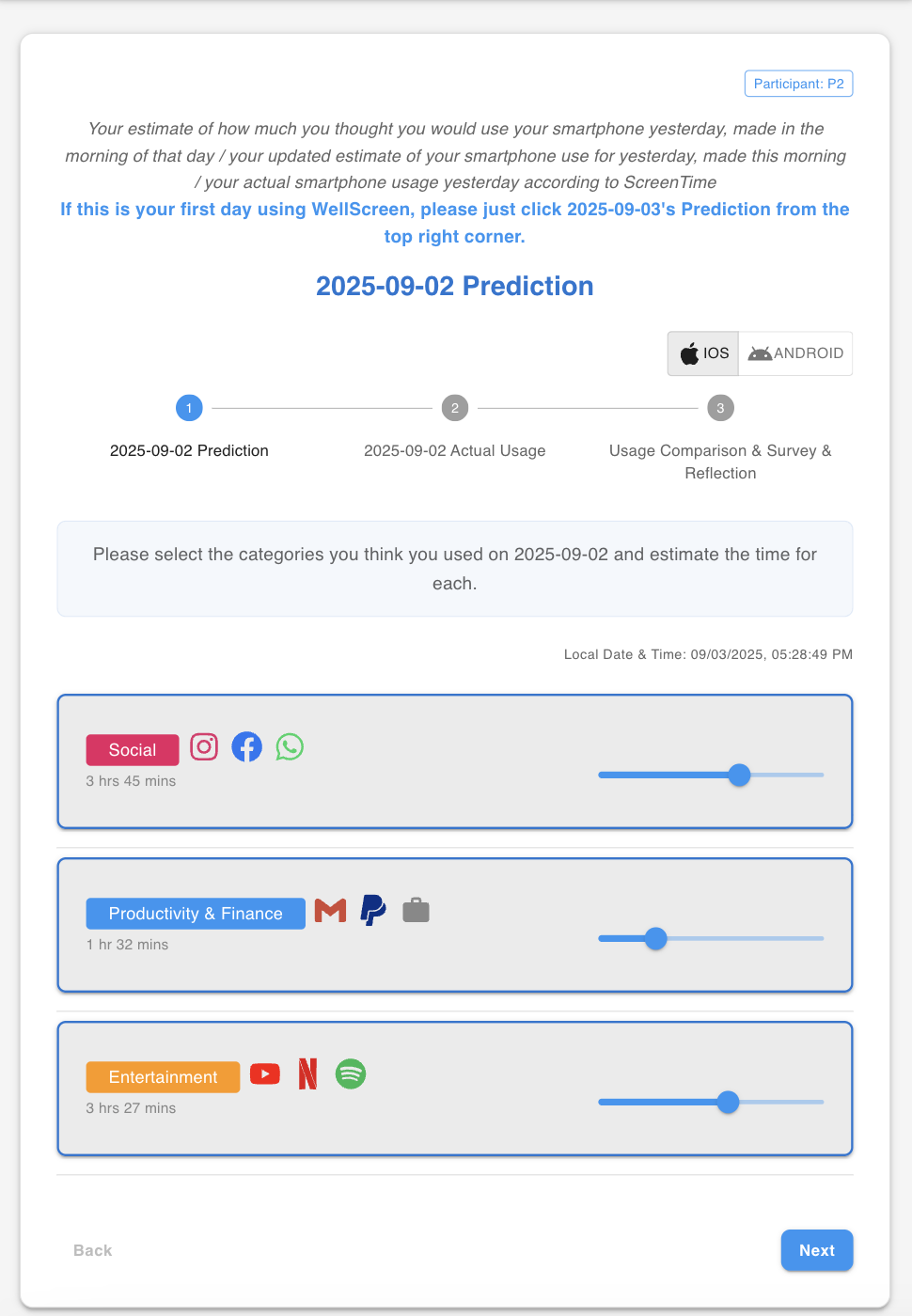}
    \caption{End-of-the-day Estimation}
    \end{subfigure}\hfill
\begin{subfigure}[b]{0.66\columnwidth}
    \centering
    \includegraphics[width=\columnwidth]{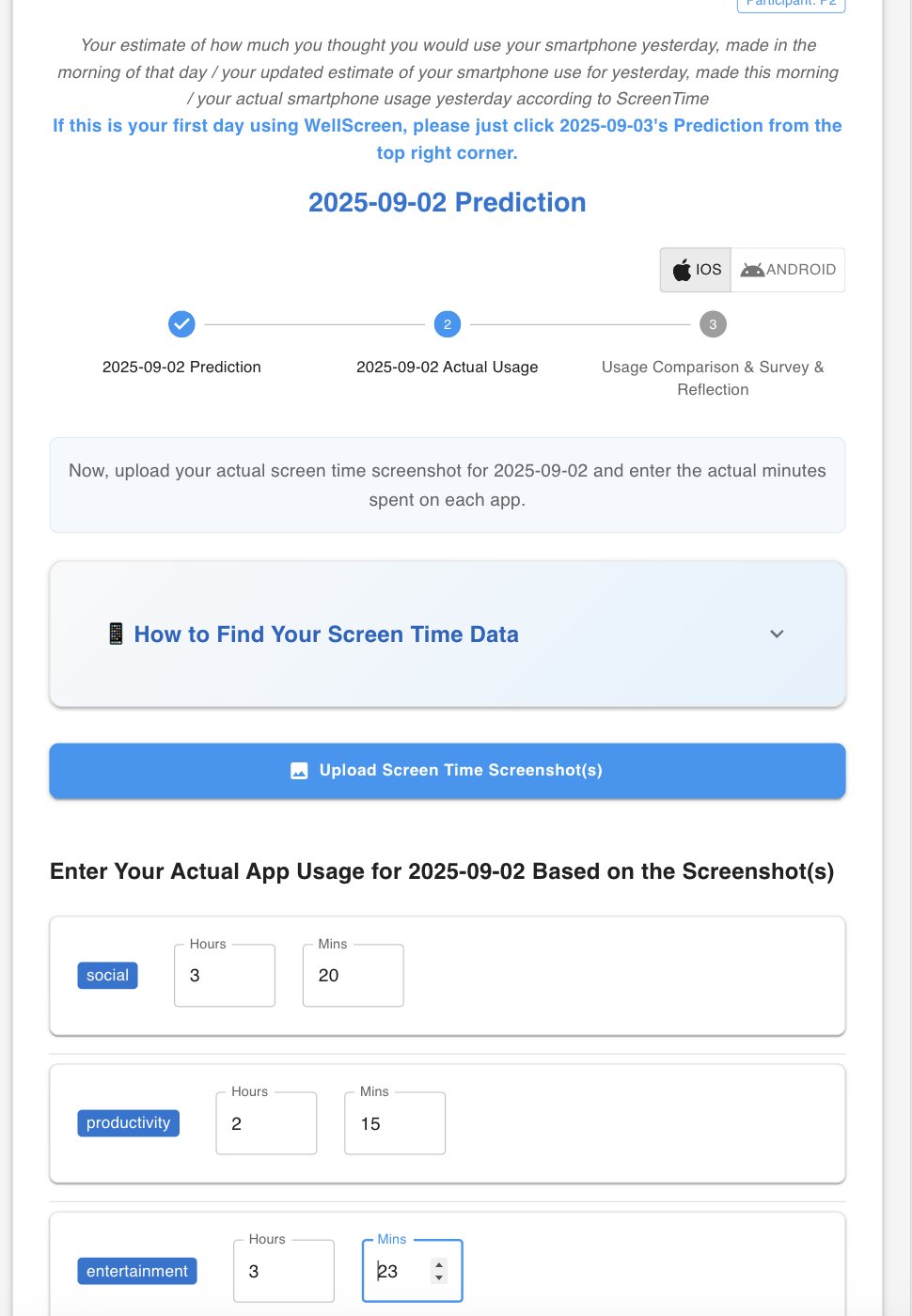}
    \caption{Actual Report}
    \end{subfigure}\hfill
 \begin{subfigure}[b]{0.66\columnwidth}
    \centering
    \includegraphics[width=\columnwidth]{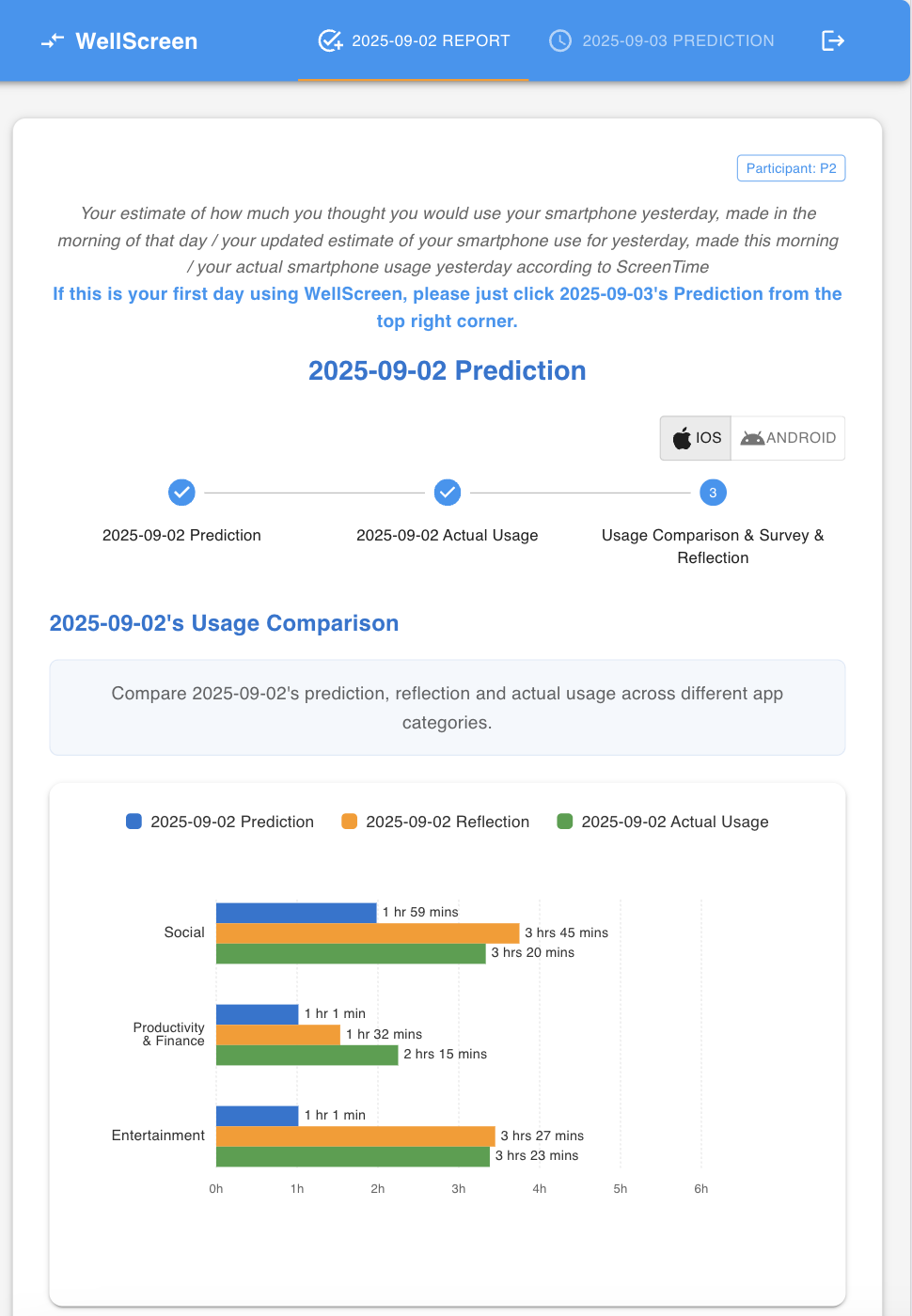}
    \caption{Visualization}
    \end{subfigure}\hfill
  \begin{subfigure}[b]{0.66\columnwidth}
    \centering
    \includegraphics[width=\columnwidth]{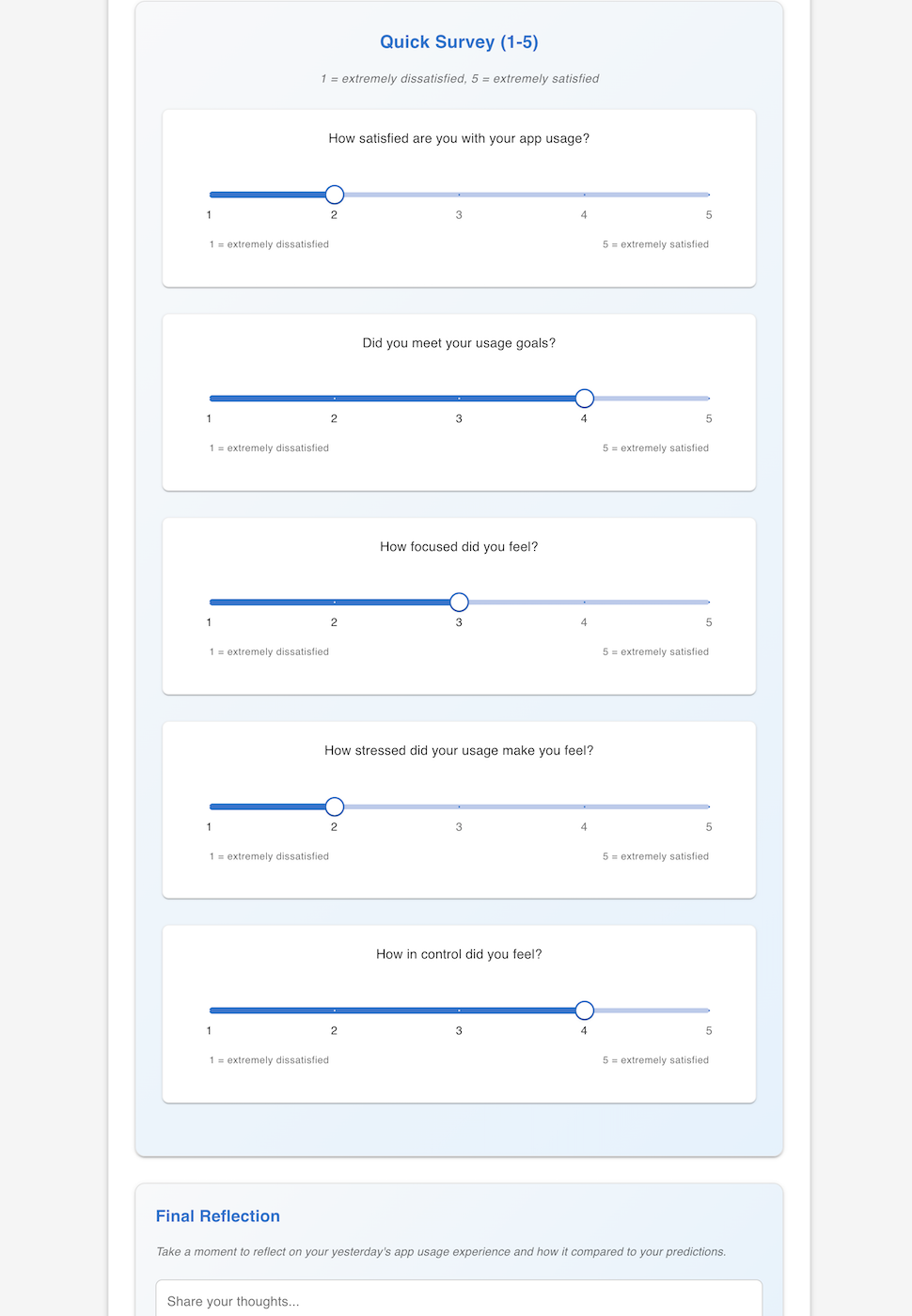}
    \caption{Reflection Survey}
    \end{subfigure}\hfill
    
    \caption{Screenshots of the \ws{} web application.} 
    \label{fig:wellscreen}
    \Description[figure]{This figure presents screenshots of the WellScreen web application interface used in the study. The interface consists of six sequential components: (a) the login page, where participants enter their assigned ID and password to begin the session; (b) the start-of-day estimation screen, where participants predict their smartphone usage duration across five activity categories using sliders; (c) the end-of-day estimation screen, which prompts users to provide their end-of-day estimates across the same categories; (d) the actual report page, where participants manually input their real usage data based on their smartphone’s screen time screenshot; (e) the visualization module, which displays a comparative bar chart of the participant’s predicted, end-of-day reflected, and actual usage; and (f) the reflection survey, which includes five Likert-style items evaluating participants’ awareness and experience with their estimations, along with a final reflection question about their overall smartphone use that day.}
\end{figure*}

We intentionally designed \ws{} to scaffold reflection through a series of lightweight interactions. As illustrated in~\autoref{fig:wellscreen}, the system guided participants through six core screens: (a) login, (b) Start-of-the-day (SoD) estimation, (c) End-of-the-day (EoD) estimation, (d) actual report, (e) visualizing the comparison of estimation, reflection, and actual report, and (f) a self-reflection survey. 
By juxtaposing SoD and EoD estimations with their actual usage, \ws{} aimed to make discrepancies visible while encouraging participants to critically evaluate their digital habits.


\subsection{Participants and Recruitment}

\begin{table*}[t!]
\centering
\sffamily
\footnotesize
   \caption{Demographic information of the 25 participants who enrolled in the study, including Age, Gender, Race, Education, and Occupation. The ``Exit'' column marks (\checkmark{}) the 20 participants who stayed the entire two-weeks duration of the study, and completed the exit survey and interview.} 
   \label{tab:participants}
\setlength{\tabcolsep}{4pt}
\begin{tabular}{lllllllllll}
\textbf{ID} & \textbf{Age (Ys)} & \textbf{Sex} & \textbf{Race} & \textbf{Education} & \textbf{Student Status} & \textbf{Device} & \textbf{Daily Use} & \textbf{Exit}  \\
\toprule
P04 &  25-35 & Male & Asian & Bachelor's degree & Master's & Android & 8-10 Hrs. & \checkmark{}\\

\rowcollight P05 & 25-35 & Female & Asian, White & Advanced degree & Master's & iOS & 5-7 Hrs. & \checkmark{}\\ 

P06 & 19-24  & Male & Asian & Some college, no degree &  Undergraduate & Android & 8-10 Hrs. & \checkmark{} \\

\rowcollight P07 & 19-24  & Female & Asian & Bachelor's degree & Recent graduate & iOS & 11-14 Hrs. & \checkmark{}\\

P08 & 19-24 & Female & Asian & Bachelor’s degree & Master's & Android & 5-7 Hrs. & \checkmark{} \\

\rowcollight P09 & 19-24  & Male & Asian & High school or equivalent & Undergraduate & iOS & 8-10 Hrs.  &  \\

P10 & 19-24  & Female & Asian & High school or equivalent & Undergraduate & iOS & 5-7 Hrs.  & \\

\rowcollight P11 & 19-24  & Female & Asian & Bachelor's degree & Undergraduate & iOS & 11-14 Hrs. &  \\

P12 & 19-24  & Male & Asian & High school or equivalent &  Undergraduate  & iOS & 8-10 Hrs.  & \checkmark{} \\

\rowcollight P13 & 25-35  & Male & Asian & Advanced degree & Master's & iOS & 5-7 Hrs.  & \checkmark{} \\

P14 & 19-24  & Female & White & Some college, no degree & Undergraduate & iOS & 8-10 Hrs.  &\\

\rowcollight P15 & 19-24  & Male & Asian, White & Some college, no deg. & Undergraduate & iOS & 5-7 Hrs.  &\checkmark{} \\

P16 & 19-24  & Female & Hispanic or Latino & Associate degree & Undergraduate & iOS & 8-10 Hrs.  & \checkmark{}\\

\rowcollight P17 & 19-24  & Female & Asian & High school or equivalent & Undergraduate & iOS & 5-7 Hrs. & \checkmark{} \\

P18 & 19-24  & Female & Asian & Some college, no degree &  Undergraduate & iOS & 5-7 Hrs.  & \checkmark{} \\

\rowcollight P19 & 19-24  & Male & White & Bachelor's degree & Master's & iOS & 5-7 Hrs.  &  \\

P20 & 19-24  & Male & American Indian/Alaska Native & Bachelor's degree & Undergraduate & iOS & 5-7 Hrs.  & \checkmark{}\\

\rowcollight P21 & 50-65  & Male & White & Advanced degree &  Some other program & Android & 14+ Hrs. &\checkmark{}\\

P22 & 19-24  & Male & Asian & High school or equivalent &  Undergraduate & iOS & 8-10 Hrs.  & \checkmark{} \\

\rowcollight P23 & 25-35  & Female & White & Advanced degree & PhD & iOS & 11-14 Hrs.  & \checkmark{} \\

P24 & 19-24  & Female & Black/African American, White & Some college, no degree & Undergraduate & Android & 5-7 Hrs.  & \checkmark{} \\

\rowcollight P25 & 19-24  & Male & White, Middle Eastern  & Associate degree &  Undergraduate & iOS & 5-7 Hrs. & \checkmark{} \\

P26 & 19-24  & Male & Asian & Bachelor's degree & Undergraduate & Android & 2-4 Hrs. &\checkmark{} \\

\rowcollight P27 & 25-35  & Male & Asian & Bachelor’s degree &  Master's & Android & 5-7 Hrs.  &\checkmark{}  \\

P28 & 25-35  & Female & Asian & Bachelor's degree & Master's & Android & 14+ Hrs. & \checkmark{} \\
\bottomrule
\end{tabular}
\Description[table]{This table presents demographic characteristics of the 25 participants in the study. The table summarizes self-reported information, including age group, gender, race/ethnicity, highest level of education, current student status, primary smartphone device type (iOS or Android), typical daily smartphone usage range, and whether the participant completed the study (Exit ✓).}
\end{table*}

\begin{table}[t]
\sffamily
\centering
\footnotesize
\caption{Descriptive statistics of self-reported individual differences of the \n{N}=25 participants enrolled in the study. The distributions include the range, mean (\n{\mu}), and standard deviation (\n{\sigma}).}
\resizebox{\columnwidth}{!}{
\begin{tabular}{lllr}
    \textbf{Covariates} & \textbf{Distribution} \\
    \cmidrule(lr){1-1} \cmidrule(lr){2-4}
    \rowcollight \multicolumn{4}{l}{\textit{Personality Trait (BFI scale)}} \\
    ~Extraversion & [1, 5], \n{\mu}=2.94, \n{\sigma}=0.83 &
    \includegraphics[width=0.45in]{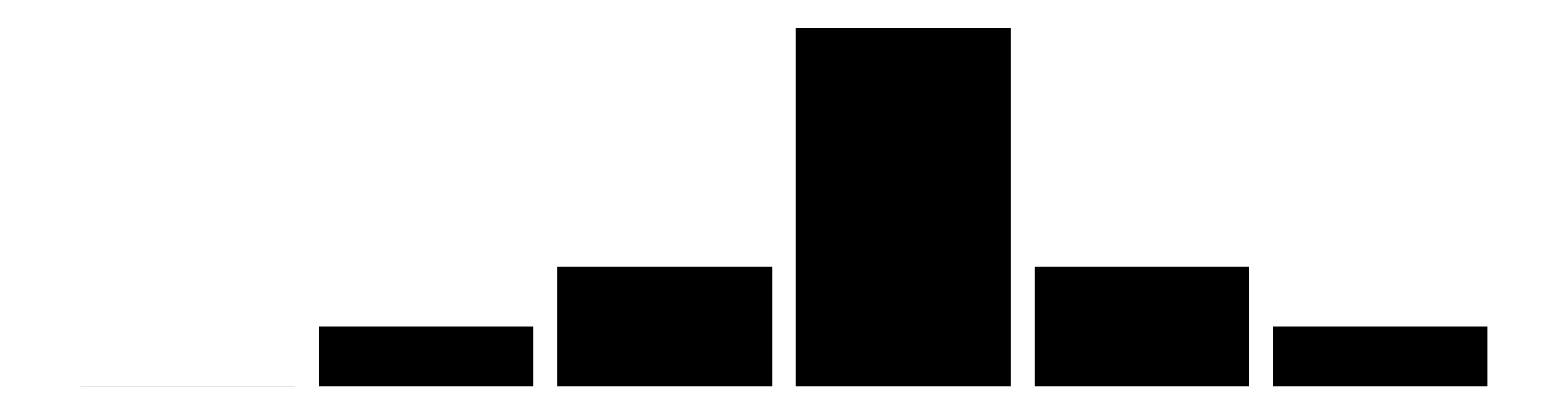} \\
    ~Agreeableness & [2.5, 4.5], \n{\mu}=3.6, \n{\sigma}=0.69 &
    \includegraphics[width=0.45in]{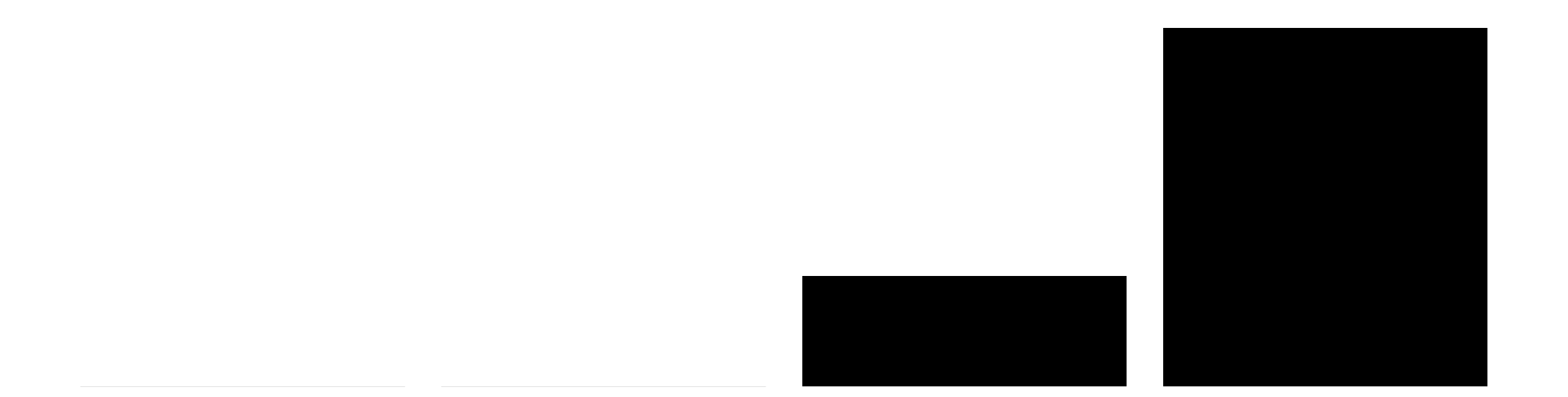} \\
    ~Conscientiousness & [1, 4.5], \n{\mu}=3.54, \n{\sigma}=0.75 &
    \includegraphics[width=0.45in]{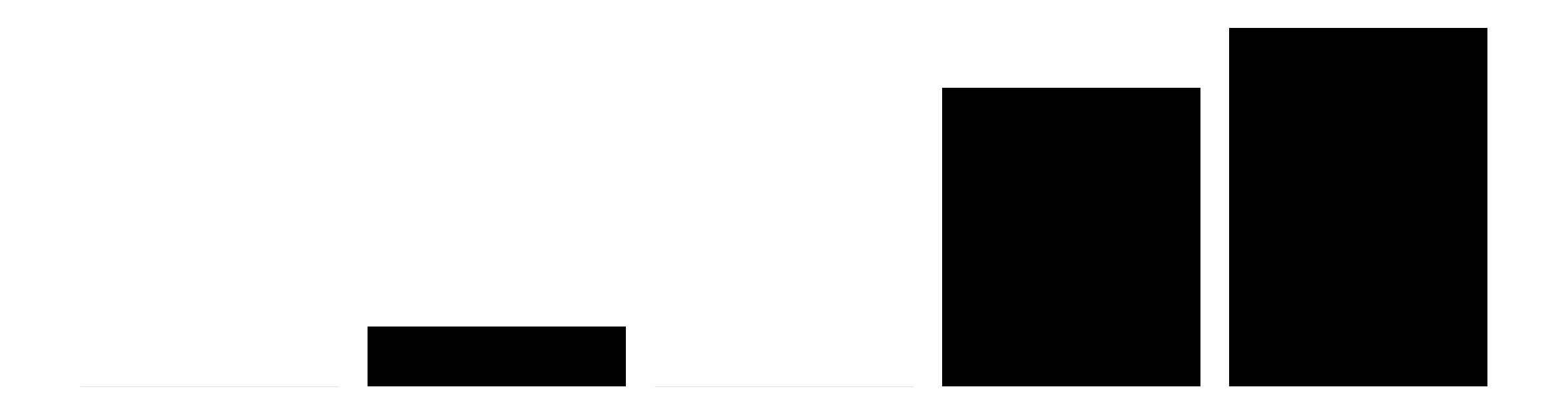}\\
    ~Neuroticism & [1.5, 5], \n{\mu}=3.22, \n{\sigma}=0.86 &
    \includegraphics[width=0.45in]{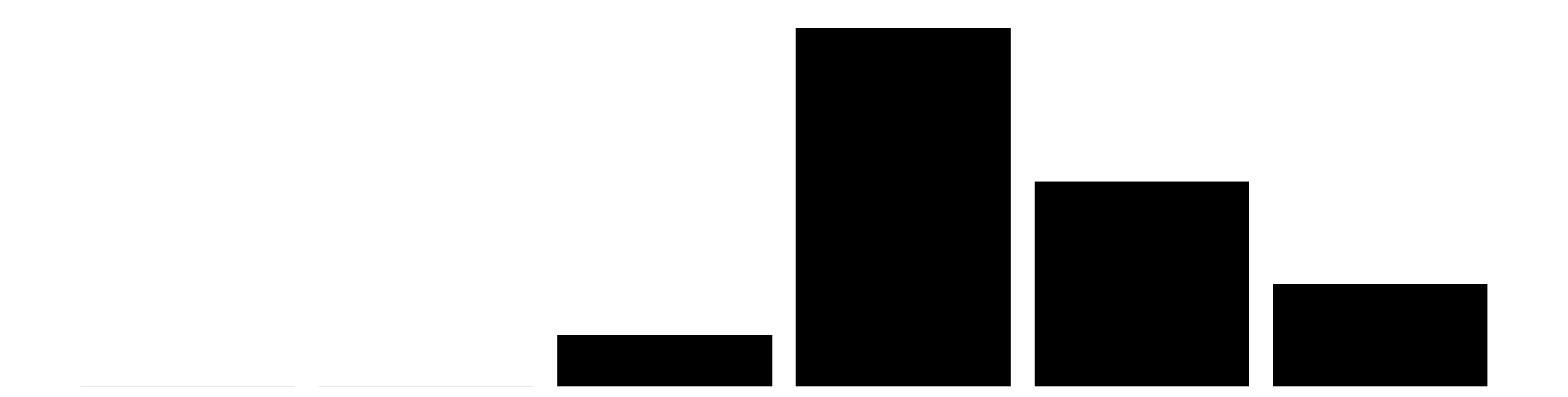} \\
   ~Openness & [1.5, 5], \n{\mu}=3.42, \n{\sigma}=0.86 &
    \includegraphics[width=0.45in]{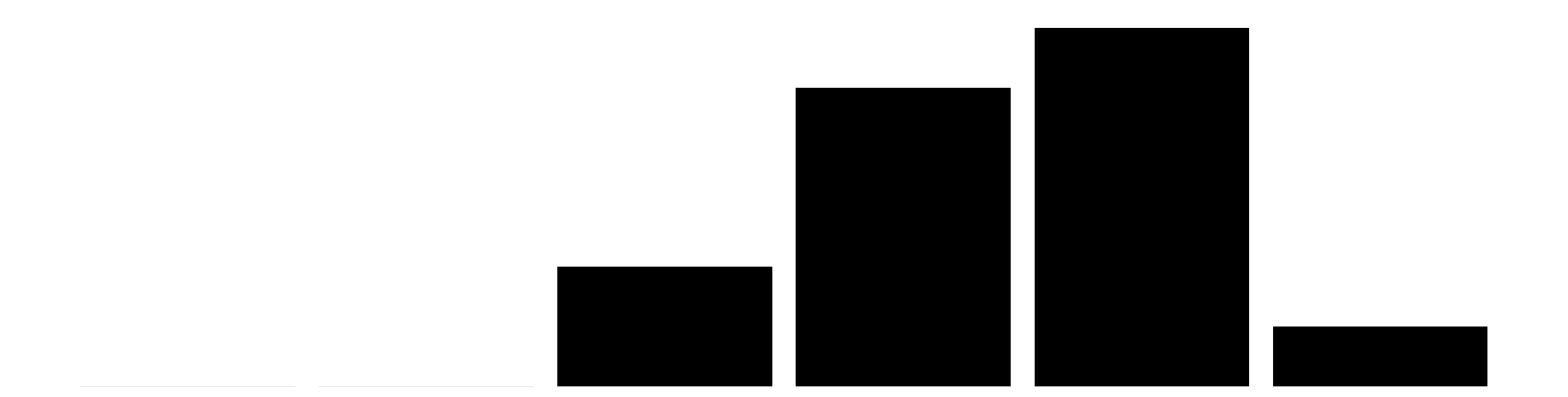} \\

    \rowcollight \multicolumn{4}{l}{\textit{Self-Regulation  and Self-Control}} \\
    ~Short Self-Regulation Questionnaire (SSRQ) & [27, 55], \n{\mu}=46.04, \n{\sigma}=6.19 &
    \includegraphics[width=0.45in]{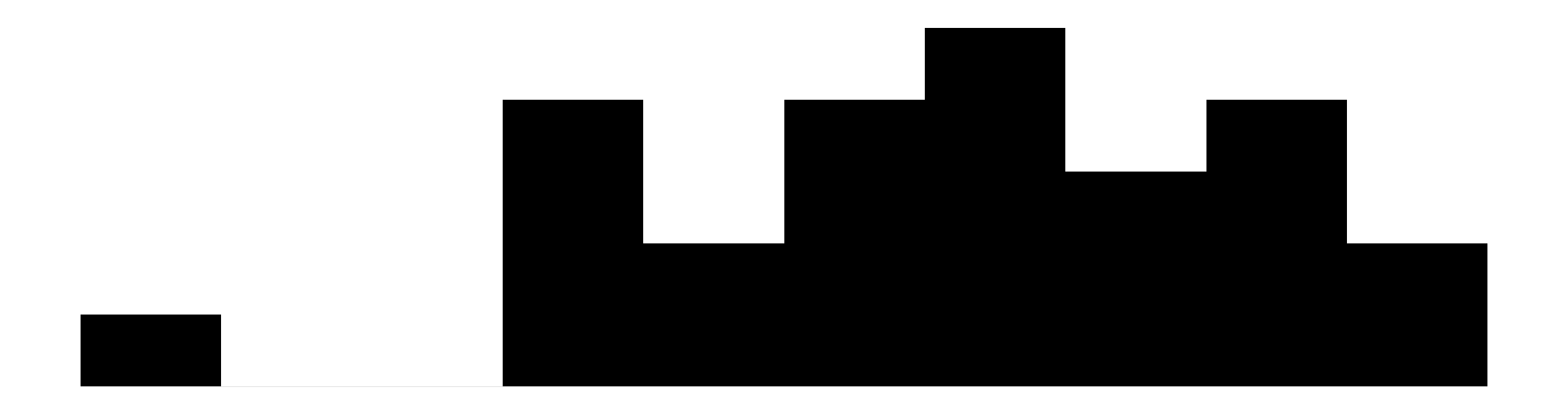} \\
    ~Basic Self-Control Scale (BSCS) & [1.7, 4.5], \n{\mu}=3.43, \n{\sigma}=0.65 &
    \includegraphics[width=0.45in]{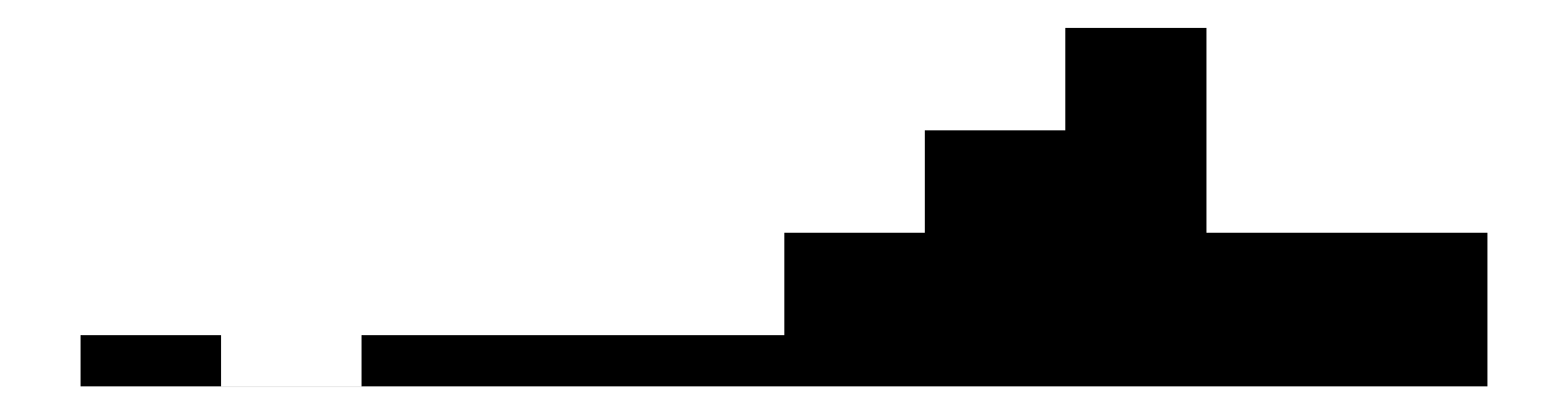} \\
    \rowcollight \multicolumn{4}{l}{\textit{Emotional Wellbeing}} \\
    ~Positive Affect (PANAS-SF) & [23, 39], \n{\mu}=30.92, \n{\sigma}=4.77 &
    \includegraphics[width=0.45in]{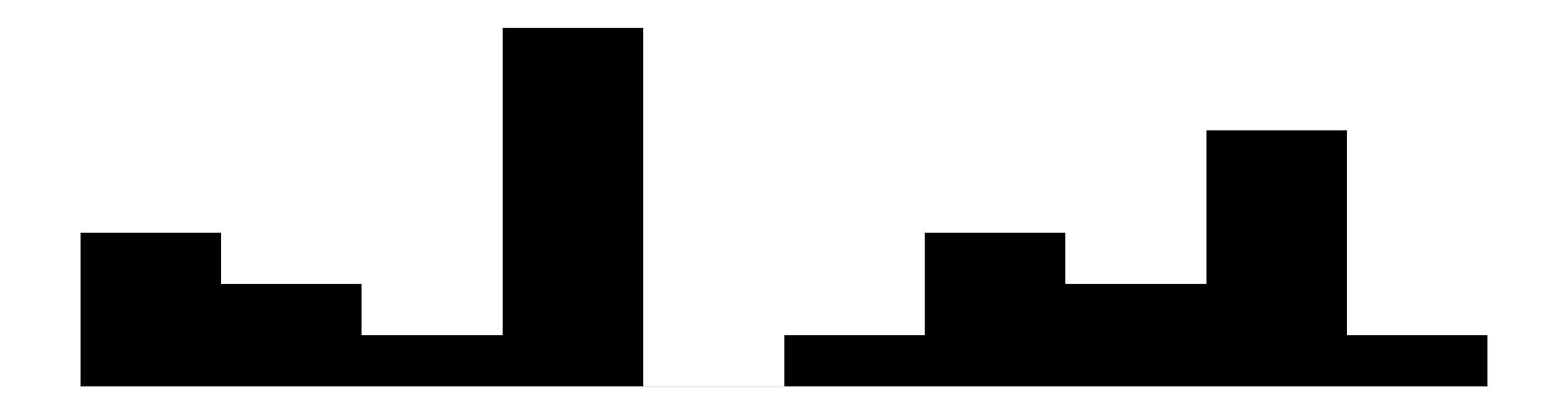} \\
    ~Negative Affect (PANAS-SF) & [13, 50], \n{\mu}=26.0, \n{\sigma}=8.53 &
    \includegraphics[width=0.45in]{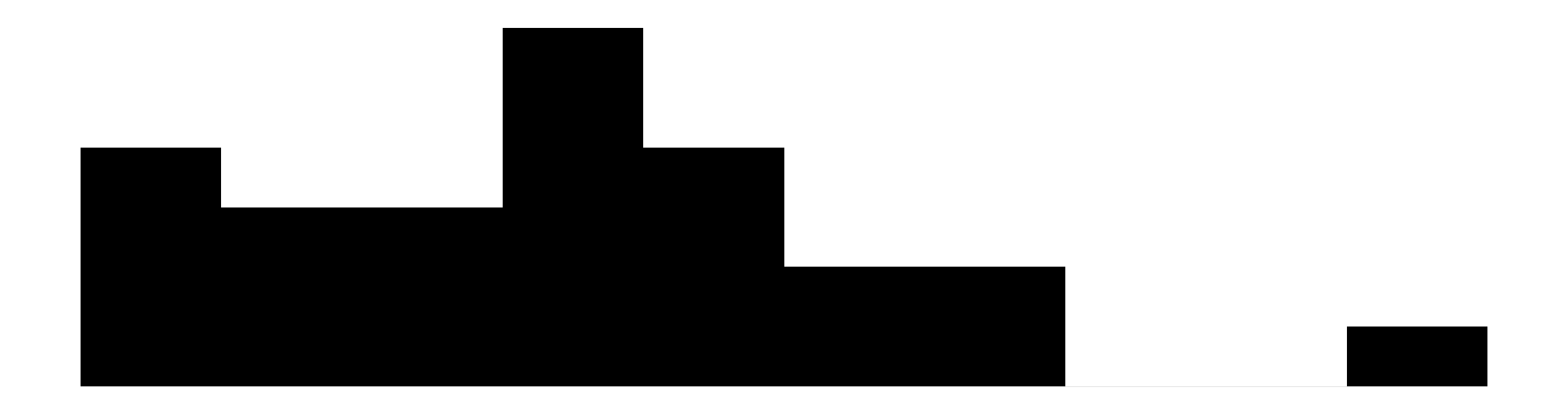} \\
    \bottomrule
\end{tabular}}
\label{table:demographics}
\Description[table]{This table summarizes the descriptive statistics of self-reported demographic and psychological covariates from the 25 participants in the study. The covariates are organized into three categories: personality traits (assessed using the BFI scale), self-regulation and self-control, and emotional wellbeing. For each measure, the table displays the observed range, mean (μ), standard deviation (σ), and a histogram illustrating the distribution across participants. Personality trait scores are calculated as item-level means on a 1 to 5 scale, while the self-regulation and emotional wellbeing measures are based on summed or averaged scores across multiple items, resulting in broader value ranges determined by their respective instruments.}
\end{table}

We recruited our participants through social media, particularly from online college communities on Reddit. Reddit is widely popular among youth and the college student demographic~\cite{statista2021reddit}, and has college-specific communities or college subreddits~\cite{saha2017stress}.
In particular, on the college subreddits of the universities where authors hold affiliations, we posted our recruitment flyer with an interest form that included a demographic survey questionnaire (age, sex, race, U.S. state) and their student status. 
Interested participants were asked to provide their institutional email IDs, enabling us to verify identities and filter out bots, duplicates, and fake responses from the shortlisted participant pool. 

Our inclusion criteria were that participants were 18 years or older, living in the US, and used digital devices (such as smartphones, laptops, tablet PCs, etc.) in their everyday lives. 
We received \num{91} responses to our interest form over a two-month period (June--Aug 2025), and we invited a subset of participants to maximize diversity and balance across answers to the students' program. This led to a final set of 25 participants who consented to participate in the study, and were invited for an interview and onboarded with the \ws{} technology probe.  
Participants were compensated using a staggered strategy: they received a \$20 USD Amazon gift card after completing the entry interview and one week of participation, and an additional \$20 USD gift card after the second week of participation and the exit interview. Participant information is available in \autoref{tab:participants}. 

\subsection{Deployment of \ws{}}

We conducted a three-phase study that included entry and exit surveys and interviews, and a 14-day deployment of our technology probe \ws{}, during which participants reported estimated and actual screen use and completed survey reflections. We describe these phases below.

\subsubsection{\edit{Choice of Survey Instruments}}
\edit{To examine how self-reflective engagement with \ws{} related to changes in self-awareness and wellbeing, and the usability and acceptability of the \ws{} tool, we employed instruments that capture both stable individual traits (covariates) and short-term psychological and usability outcomes.}

\edit{For this purpose, we assessed personality traits using the BFI-10 scale~\cite{soto2017next}---prior work has associated personality traits with technology use tendencies, attentional control, and emotional reactivity~\cite{das2019multisensor}. in addition, the Short Self-Regulation Questionnaire (SSRQ) and the Brief Self-Control Scale (BSCS) capture complementary aspects of regulatory capacity: SSRQ reflects goal management and self-monitoring abilities relevant to estimation accuracy and reflective practice~\cite{carey2004psychometric}, whereas BSCS measures trait impulse control linked to behavioral consistency and the ability to align intentions with actual technology use~\cite{tangney2018high}. 
To assess wellbeing outcomes, we used the PANAS-SF to measure changes in positive and negative affect following the \ws{} deployment. 
In the exit survey, we additionally included the System Usability Scale (SUS)~\cite{brooke1996sus} and the Intervention Appropriateness Measure (IAM)~\cite{weiner2017psychometric} to evaluate participants’ perceived usability of WellScreen and its suitability as a reflection-based wellbeing intervention.}

\edit{Together, these measures provided both trait-level covariates (personality, self-regulation, self-control) and outcome-level assessments (affect, usability, and appropriateness) that supported our modeling and interpretation of awareness gaps, reflective engagement, and wellbeing change across the study.}

\subsubsection{Phase 1: Entry Survey and Interview}

During enrollment, participants responded to an entry survey, consisting of validated trait-based questionnaires across--- 1) personality traits through BFI-10~\cite{soto2017next}, 2) self-regulation skills with respect to digital use adapted from the Short Self-Regulation Survey Questionnaire (SSRQ)~\cite{carey2004psychometric}, 3) self-control traits through the Brief Self-Control scale~\cite{tangney2018high}, and 4) emotional wellbeing traits in terms of affect through the PANAS-SF scale~\cite{watson1988development,mackinnon1999short}. Summary statistics are presented in \autoref{table:demographics}.
In addition, participants responded to questions on the typical daily use of digital devices. 


During enrollment, we also conducted semi-structured remote interviews via Teams video calls. These interviews lasted between 30 and 45 minutes and were conducted in English. We recorded these interviews and transcribed them soon after they were completed. During these interviews, we asked our participants questions about their typical daily digital device usage practices, their understandings of the concept of digital wellbeing, and any prior experience with tracking their smartphone usage behaviors. We ended these interviews with a demonstration of \ws{}.

\subsubsection{Phase 2: Digital Activity Logging}
We asked participants to use \ws{} for 14 days, starting the day after their entry interview. We instructed them to log their SoD estimation, EoD estimation, and actual reports each morning for 14 consecutive days. On the first day, participants began by making a SoD estimation of their expected smartphone usage for that day. \edit{The estimations were based on the default ``App Categories'' available on Android and iOS devices to minimize participants' cognitive load in engaging with \ws{}. We included five categories that were most likely to be used and also were consistently named across platforms: Social, Productivity \& Finance, Shopping \& Food, Entertainment, and Creativity. We included an ``Other'' category to capture miscellaneous usage.} From the second day onward, the routine expanded to include their estimation of the accuracy of their previous day’s predictions (through their EoD estimation), followed by the recording of actual usage values transcribed from a screenshot of their phone’s usage summary. Once this information was submitted, \ws{} generated a visualization that displayed side-by-side comparisons of the SoD estimation, EoD estimation, and actual usage across multiple app categories. 
Building on this visualization, participants then completed a short survey and wrote a brief reflection on their usage patterns of the previous day, before concluding the cycle by making a new SoD estimation for the current day. This daily process was repeated for the duration of the study, creating an iterative loop of SoD and EoD estimations, and comparison that juxtaposed participants' estimations with their own device use.

To support consistency, we sent a reminder at 8AM and, if necessary, a follow-up reminder at 10AM in participants' local time zone. 
We chose this once-daily interaction schedule to accommodate variability in participants' natural sleep and usage routines while ensuring that EoD estimations captured the entirety of the previous day. We used the screenshots to verify the manually entered actual reports. 
We strongly encouraged participants to log their data every day for 14 consecutive days. 
\edit{Out of the 25 participants who enrolled in the study, 20 participants completed 14 days of data logging and participated in the exit interview and survey (see \autoref{table:demographics}). 
Among these 20 participants, 9 completed this consecutively over 14 days, while the remaining 11 completed the study within $16.2 \pm 1.23$ days of participation.}

\subsubsection{Phase 3: Exit Survey and Interview}
We scheduled exit interviews with 20 participants who completed 14 days of self-reporting. We sent the participants an exit survey containing validated trait-based questionnaires examining: 1) emotional wellbeing traits in terms of affect through the PANAS-SF scale~\cite{watson1988development,mackinnon1999short}, 2) self-regulation skills with respect to digital use adapted from the Short Self-Regulation Survey Questionnaire (SSRQ)~\cite{carey2004psychometric}, and 3) self-control traits through the Brief Self-Control scale~\cite{tangney2018high}. We additionally included a system usability scale (SUS)~\cite{brooke1996sus} and intervention appropriateness measure (IAM)~\cite{weiner2017psychometric}. We then interviewed these participants via Teams video calls. 
These interviews ran between 30 and 45 minutes and were audio-recorded and transcribed. 
In these exit interviews, we asked questions about their experiences using \ws{}, reflections on their estimated--actual gaps (E--A gap), and reflections on any behavior change on their part: either during their participation or going forward. 



\subsection{Data Analysis}

\subsubsection{Quantitative Analysis} 
Our quantitative analysis was conducted to measure and evaluate participants' self-estimates of their smartphone usage. The data for this section was sourced from: 1) the daily self-reports by \ws{} participants, which captured both estimated and actual smartphone usage, 2) daily self-reports of wellbeing surveys, and 3) the entry and exit surveys, which provided self-reported emotional wellbeing and psychological traits data. 
To quantify the E--A gap, we employed three key metrics: Relative difference ($\Delta$\%), Symmetric Mean Absolute Percentage Error (SMAPE), and paired $t$-tests. 
Relative difference ($\Delta$\%) was calculated to show the direction and magnitude of the \ea{}. Negative value of $\Delta$\% indicates under-estimation (estimated use lower than the actual use) and positive values indicate over-estimation (estimated use higher than actual use). 
Paired $t$-tests provided the statistical significance of the differences between estimated and actual usage.

In addition, we conducted linear mixed effects regression models to understand the relationship between individual differences with SoD and EoD \ea{}s. 
We also built linear mixed effects regression models to understand the relationship between \ea{}s and daily self-reported wellbeing measures, including satisfaction, goal adherence, self-control, focus, and stress. 
These models controlled for demographic variables and individual difference measures (e.g., personality traits, self-control, and self-regulation) collected in the entry survey. Finally, to assess changes in psychological outcomes, we compared exit and entry surveys using paired $t$-tests and quantified the differences with effect sizes (Cohen’s $d$).


\subsubsection{Qualitative Analysis} 
Following the interviews, all recordings were transcribed using the default transcription feature in Teams. The transcripts were then anonymized by redacting any identifiable data, and the complete data---including transcripts and interview notes---were used as corpus for analysis. Our qualitative analysis was conducted using a reflexive thematic analysis~\cite{braun2019reflecting}. All authors were involved in an interactive and iterative process of open coding. Two co-authors led the initial coding of the raw interview transcripts, while the others provided feedback and guidance during hybrid co-working sessions. The two authors also conducted the high-level thematic analysis to identify themes from the low-level codes. This collaborative approach was crucial for ensuring the coherence of our analysis.  

During the iterative process, we carefully reviewed and refined the emergent themes. This involved merging related codes into broader categories, separating overlapping themes into distinct ones, and discarding themes that were not directly relevant to our research questions. 
From an initial 581 (from entry interview) + 487 (from exit interview) codes, we iteratively refined our codes into 5 high-level themes 
that aligned directly with our research questions. 





\subsection{Privacy, Ethics, and Positionality}
Our study was reviewed and approved by the Institutional Review Boards (IRB) at our universities.
Given the sensitive nature of personal wellbeing and technology use that surfaced during the deployment study and interviews, we implemented strict privacy and ethical safeguards.
Our datasets were stored in secure databases and protected server environments.
Before analyzing our data, we removed all personally identifiable information and paraphrased quotes in the paper to reduce traceability while maintaining the context for readers. Our interdisciplinary team comprises researchers of diverse gender, racial, and cultural backgrounds, including people of color and immigrants.
The team brings expertise in HCI, CSCW, human-centered design, and AI ethics. 
The authors have conducted prior studies on digital wellbeing and technology use, and as individuals with lived experiences of technology in both academic and everyday contexts, our interpretations are also informed by these perspectives.
While we have taken the utmost care to faithfully synthesize participants' viewpoints, we acknowledge that our interpretations are situated and shaped by our disciplinary training, professional backgrounds, and personal experiences.
\section{RQ1: Estimated--Actual Gap in Digital Use}


This section focuses on calculating and understanding our participants' estimated--actual gap (E--A gap) as a dimension of self-awareness that could promote their digital wellbeing. In doing so, we address our \textbf{RQ1:} \textit{How do users' estimations of digital use compare with their actual use, and what factors explain the disparities?}


\subsection{Quantifying the Estimated--Actual Gap}

\autoref{tab:cross-sectional} summarizes the participants' E--A gap in smartphone use differentiated by app categories, quantified using relative difference ($\Delta$\%), symmetric mean absolute percentage error (SMAPE), and paired $t$-tests for statistical significance. Negative values of relative difference ($\Delta$\%) indicate under-estimation (estimated lower than actual use), whereas positive values indicate over-estimation (estimated higher than actual use).  

At an aggregate level, participants' estimated and actual use did not significantly differ. However, when disaggregated by app category, we observe interesting patterns: participants significantly \textit{over-estimated} \ent{} app use by 11.32\%, while \textit{under-estimating} \pro{} use by 16.77\% and \soc{} app use by 9.49\%. 
The contrasting trends between \ent{} and \pro{} may be connected: participants could be underestimating time spent on work- or productivity-related activities, or their professional obligations may limit opportunities for engaging with entertainment apps. Similarly, the under-estimation of \soc{} use is notable, as it diverges from prior findings that individuals tend to over-estimate their Facebook use~\cite{ernala2020well}.

\subsection{Factors Explaining Estimated--Actual Gap}
Next, we examined how participants' individual-level factors associated with with their E--A gaps. For this purpose, we employed linear mixed-effects regression models, which capture both longitudinal and cross-sectional variation by accounting for within-person changes over time as well as between-person differences across participants.
We built separate models for each app category (\crea{}, \ent{}, \pro{}, \shop{}, \soc{}) and one for the aggregated E--A gap. For independent variables, we included individual-level factors (demographics, psychological traits, smartphone type) and day-level controls (study day and weekend/weekday indicator). 
We conducted this analysis separately for start-of-the-day (SoD) E--A gaps and for end-of-the-day (EoD) E--A gaps.

The regression results are reported in \autoref{table:sodgap_regression} (SoD E--A gap) and \autoref{table:eodgap_regression} (EoD E--A gap). A majority of independent variables do not show statistical significance, suggesting that misestimations in app use may not systematically follow specific demographic or contextual patterns. 
However, we note that self-control shows statistical significance for a number of categories. 

In the SoD E--A gap models, self-control was negatively associated with creativity ($\beta$=–7.81), productivity ($\beta$=–7.24), and the aggregated gap ($\beta$=–16.93). Similarly, in the EoD models, self-control was negatively associated with creativity ($\beta$=–3.75), productivity ($\beta$=–4.15), and the aggregated gap ($\beta$=–11.04). 
These results indicate that individuals with higher self-control tended to have smaller E--A gaps, suggesting that self-regulatory capacity may be linked to more accurate awareness of digital use.

These quantitative findings are supported by qualitative data from our interviews, where participants offered insights into some specific factors influencing their estimation accuracy. P17, for example, noted
\textit{``a disconnect between what I think I am doing and what I am actually doing\dots Seeing my screen time has made me aware of this mismatch.''} This misestimation was often driven by situational factors, such as travel, which caused usage to unexpectedly spike, as P4 mentioned \textit{``when I was traveling\dots that is why I was on my phone quite a bit, so that sort of agrees with the usage quite a bit.''} We found that providing participants with quantified and visualized E--A gaps facilitated critical self-reflection.
Participants often experienced an initial shock or surprise at their high screen time, as P9 described, \textit{``I often have an initial shock when I see my screen time numbers, thinking, `Did I really use it that much?' but then I realize it makes sense. I reflect on my day and think, `Oh, yeah, I was on my phone from this time to this time, and on my computer during those hours.' ''} Additionally, we suspect that some participants might have chosen not to log data for a given day if they anticipated irregularities in their schedules. P23, for example, \textit{``skipped a day [of self-reporting]\dots because I had meetings and I completely forgot about the app.''} 
This findings reveal that our participants' E--A gap for screen time may primarily have been due to a lack of situational awareness and a habitual tendency to underestimate usage.

\begin{table*}[t!]
\centering
\sffamily
\footnotesize
   \caption{\textbf{Estimated-Actual (E--A) Gap:} Comparison of per-day actual app usage (in minutes) with start-of-the-day (SoD) estimations and end-of-the-day (EoD) estimations for all participants. The columns include mean occurrences, relative differences ($\Delta$ \%), and paired $t$-tests with respect to actual reports. \edit{$p$-values are reported after FDR correction.} (* $p$<0.05, ** $p$<0.01, *** $p$<0.001). The length of horizontal bars represents the magnitude of $\Delta$ \%, where \textcolor{improveCol}{\textbf{PURPLE}} bars indicate an \textcolor{improveCol}{under-estimation} and \textcolor{worsenCol}{\textbf{GREEN}} bars indicate an \textcolor{worsenCol}{over-estimation} of app usage. }   \label{tab:cross-sectional}
\setlength{\tabcolsep}{2pt}
\begin{tabular}{lrrrclrlrrclrl}
 & \textbf{Actual} & \multicolumn{6}{c}{\textbf{SoD Estimation}}  & \multicolumn{6}{c}{\textbf{EoD Estimation}} \\
\cmidrule(lr){2-2} \cmidrule(lr){3-8} \cmidrule(lr){9-14}
\textbf{Category} & \textbf{Mean} & \textbf{Mean} & \multicolumn{3}{c}{\textbf{\ea{}} (\textbf{$\Delta$\%})} &  \multicolumn{2}{c}{\textbf{$t$-test}} & \textbf{Mean} & \multicolumn{3}{c}{\textbf{\ea{}} (\textbf{$\Delta$\%}})  &  \multicolumn{2}{c}{\textbf{$t$-test}}\\
\cmidrule(lr){1-1} \cmidrule(lr){2-2} \cmidrule(lr){3-3} \cmidrule(lr){4-6}\cmidrule(lr){7-8}\cmidrule(lr){9-9}\cmidrule(lr){10-12} \cmidrule(lr){13-14}
Creativity & 21.42 & 23.61 &  & 9.27 & \worbar{0.92} &  0.59 &  & 21.05 & \impbar{.178} & -1.78 &  &  -0.14 & \\
\rowcollight Entertainment & 104.64 & 118 & & 11.32 & \worbar{1.132}  &  2.32 & * & 112.27 & & 6.79 &\worbar{.679}  & 1.70 & \\
Productivity & 69.04 & 59.13 &\impbar{1.677} & -16.77 &  &  -2.14 & ** & 62.13 &\impbar{1.112} & -11.12 &  &  -1.97 & *\\
\rowcollight Shopping & 21.40 & 26.35 &  & 18.80 & \worbar{1.88} &   1.54 &  & 29.36 &  & 27.13 &\worbar{2.713} &  3.24 & **\\
\soc{} & 165.02 & 150.72 &\impbar{0.949} & -9.49 &  &  -2.54 & * & 167.00 &  & 1.19 & \worbar{.119} &  0.41 & \\
\rowcollight Aggregate & 381.53 & 377.8 & \impbar{0.099} & 0.99 &  & -0.28 &  & 391.82 &  & 2.63 & \worbar{.263} &  1.12 & \\
\bottomrule
\end{tabular}
\Description[table]{This table presents a comparison of daily actual app usage and participants’ estimations reported at the start-of-day (SoD) and end-of-day (EoD) across five usage categories: creativity, entertainment, productivity, shopping, and social, along with an aggregated total. For each estimation point, the table displays the mean actual and estimated usage, the relative difference between estimation and actual usage ($\Delta$\%), symmetric mean absolute percentage error (SMAPE), and the results of paired $t$-tests assessing differences from actual values. Horizontal bars visually indicate the magnitude and direction of $\Delta$\%, with purple representing under-estimation and green representing over-estimation. Statistically significant differences based on $t$-tests are indicated alongside each estimation gap. Aggregated values for all categories are included in the final row.}
\end{table*}

\begin{table*}[t]
\sffamily
\centering
\footnotesize
\caption{\textbf{SoD E--A Gap:} Linear mixed effects regression models revealing the relationship between individual differences with \textbf{SoD estimation--actual (E-A) gap}; (\textbf{Bolded} values are for \textsuperscript{.} \n{p}\textless{}0.1, * \n{p}\textless{}0.05, ** \n{p}\textless0.01, *** \n{p}\textless{}0.001).}
\begin{tabular}{l r@{}l r@{}l r@{}l r@{}l r@{}l r@{}l}
    \textbf{Dep. Variables} $\rightarrow{}$ & \multicolumn{2}{c}{\textbf{Creativity}} & \multicolumn{2}{c}{\textbf{Entertainment}} & \multicolumn{2}{c}{\textbf{Productivity}} & \multicolumn{2}{c}{\textbf{Shopping}} & \multicolumn{2}{c}{\textbf{Social}}& \multicolumn{2}{c}{\textbf{Aggregated}}\\ 
    \toprule
    \textbf{Indep. Variables} $\downarrow{}$ & \textbf{Coeff.} & & \textbf{Coeff.} &  &\textbf{Coeff.} &  &\textbf{Coeff.} &  &\textbf{Coeff.} & & \textbf{Coeff.} &  \\  
    \cmidrule(lr){1-1}\cmidrule(lr){2-3}\cmidrule(lr){4-5}\cmidrule(lr){6-7}\cmidrule(lr){8-9}\cmidrule(lr){10-11}\cmidrule(lr){12-13} 
~Intercept & 225.10 &  & 371.90 &  & 173.90 &  & 34.57 &  & -334.07 &  & 471.41&\\
~Smartphone: IOS & 28.68 &  & -37.37 &  & 46.46 &  & 1.54 &  & -10.84 &  & 28.47 &\\
~Weekend &  -7.83 &  & 4.61 &  & 4.35 &  & -0.50 &  & 7.31 &  & 7.93 & \\
~Num. Study Days & -0.77 &  & -0.12 &  & 0.31 &  & \textbf{-1.69} & \textbf{**} & \textbf{2.36} & \textbf{*} & 0.08 & \\
\rowcollight \multicolumn{12}{l}{\textit{Demographics}}\\
~Age: 25-35 & 3.88 &  & -2.59 &  & 42.04 &  & 1.21 &  & 13.03 &  & 57.58& \\
~Age: 50-65 & -21.49 &  & -168.89 &  & 163.33 &  & -23.39 &  & 89.16 &  & 38.72& \\
~Sex: Male & \textbf{28.48} & \textbf{\textsuperscript{.}} & 45.40 &  & -3.67 &  & 7.79 &  & -29.83 &  & 48.18& \\
~Race: Asian & 34.56 &  & 25.32 &  & \textbf{-128.57} & \textbf{**} & 7.47 &  & \textbf{-103.65} & \textbf{\textsuperscript{.}} & -164.87& \\
~Race: Black/Af. Am. & \textbf{87.07} & \textbf{\textsuperscript{.}} & 69.29 &  & -74.69 &  & 22.95 &  & -117.27 &  & -12.65& \\
~Race: Hisp./Latino & -30.13 &  & -86.86 &  & -26.68 &  & -8.86 &  & 60.72 &  & -91.81& \\
~Race: White & 16.97 &  & 47.21 &  & \textbf{-168.92} & \textbf{***} & -0.27 &  & -79.99 &  & -185.00 & \\
~Education: Associate & 1.03 &  & 33.27 &  & -50.42 &  & -3.99 &  & -52.04 &  & -72.15 & \\
~Education: Bachelor's & 1.03 &  & 33.27 &  & -50.42 &  & -3.99 &  & -52.04 &  & -72.15 & \\
~Education: HS/Diploma & 24.25 &  & 0.52 &  & 80.35 &  & 9.85 &  & 51.63 &  & 166.59 & \\
\rowcollight \multicolumn{12}{l}{\textit{Psychological Traits}}\\
~BFI: Extraversion & -2.96 &  & -6.38 &  & -11.70 &  & -3.49 &  & -11.98 &  & -36.52& \\
~BFI: Agreeableness & -7.68 &  & -25.97 &  & 11.04 &  & 5.66 &  & 27.89 &  & 10.94& \\
~BFI: Conscientiousness & 21.64 &  & 40.26 &  & 17.45 &  & 5.64 &  & 30.53 &  & 115.52& \\
~BFI: Neuroticism & -16.44 &  & 4.15 &  & -9.98 &  & 4.06 &  & \textbf{42.80} & \textbf{***} & 24.60& \\
~BFI: Openness & 7.26 &  & -32.06 &  & 1.75 &  & 0.62 &  & -6.58 &  & -29.01 & \\
\hdashline
~SSRQ: Self-Regulation & 6.80 &  & -57.75 &  & 34.22 &  & 4.73 &  & 57.30 &  & 45.30& \\
\hdashline
~BSCS: Self-Control & \textbf{-7.81} & \textbf{***} & -2.16 &  & \textbf{-7.24} & \textbf{***} & -1.73 &  & 2.00 &  & \textbf{-16.93}& \textbf{**} \\
\hdashline
~PANAS: P. Affect & 1.80 &  & 1.25 &  & 1.58 &  & -0.55 &  & \textbf{-6.48} & \textbf{*} & -2.40& \\
~PANAS: N. Affect &  -1.07 &  & -3.13 &  & -0.46 &  & -0.08 &  & 1.72 &  & -3.01& \\
\rowcollight & \multicolumn{2}{l}{\n{R^2}=\textbf{0.24***}} &\multicolumn{2}{l}{\n{R^2}=0.10} & \multicolumn{2}{l}{\n{R^2}=\textbf{0.16**}} & \multicolumn{2}{l}{\n{R^2}=\textbf{0.08***}} &\multicolumn{2}{l}{\n{R^2}=\textbf{0.23***}} & \multicolumn{2}{l}{\n{R^2}=\textbf{0.12*}} \\
    \bottomrule
    \end{tabular}
\label{table:sodgap_regression}
\Description[table]{This table summarizes linear mixed effects regression models examining how individual differences relate to the start-of-day estimation–actual (E–A) gap in app usage across six domains: Creativity, Entertainment, Productivity, Shopping, Social, and Aggregated. Independent variables include device type, number of study days, demographic characteristics (e.g., age, sex, race, education), and psychological traits (e.g., Big Five, self-regulation, self-control, affect scores). Each cell reports the regression coefficient and p-value, with bolded values indicating statistically significant associations. R-squared (R²) values at the bottom indicate the proportion of variance in the E–A gap explained by the model for each app domain.}
\end{table*}

\begin{table*}[t]
\sffamily
\centering
\footnotesize
\caption{\textbf{EoD E--A Gap: } Linear mixed effects regression models revealing the relationship between individual differences with \textbf{EoD estimation--actual (E-A) gap}; (\textbf{Bolded} values are for \textsuperscript{.} \n{p}\textless{}0.1, * \n{p}\textless{}0.05, ** \n{p}\textless0.01, *** \n{p}\textless{}0.001).}
\begin{tabular}{l r@{}l r@{}l r@{}l r@{}l r@{}l r@{}l}
    \textbf{Dep. Variables} $\rightarrow{}$ & \multicolumn{2}{c}{\textbf{Creativity}} & \multicolumn{2}{c}{\textbf{Entertainment}} & \multicolumn{2}{c}{\textbf{Productivity}} & \multicolumn{2}{c}{\textbf{Shopping}} & \multicolumn{2}{c}{\textbf{Social}}& \multicolumn{2}{c}{\textbf{Aggregated}}\\ 
    \toprule
    \textbf{Indep. Variables} $\downarrow{}$ & \textbf{Coeff.} & & \textbf{Coeff.} &  &\textbf{Coeff.} &  &\textbf{Coeff.} &  &\textbf{Coeff.} & & \textbf{Coeff.} &  \\  
    \cmidrule(lr){1-1}\cmidrule(lr){2-3}\cmidrule(lr){4-5}\cmidrule(lr){6-7}\cmidrule(lr){8-9}\cmidrule(lr){10-11}\cmidrule(lr){12-13} 
~Intercept & 59.12 &  & 39.84 &  & -26.17 &  & 279.23 &  & -185.20 &  & 166.81 & \\
~Smartphone: IOS & 28.68 &  & -37.37 &  & 46.46 &  & 1.54 &  & -10.84 &  & 28.47 &\\
~Weekend &  -3.45 &  & -1.15 &  & 5.74 &  & -1.54 &  & 6.40 &  & 5.99 & \\
~Study Day Num. & -0.64 &  & -0.43 &  & 0.02 &  & 0.11 &  & \textbf{2.08} & \textbf{**} & 1.13 & \\
\rowcollight \multicolumn{12}{l}{\textit{Demographics}}\\
~Age: 25-35 & 21.11 &  & 34.62 &  & 59.53 &  & -42.99 &  & -30.13 &  & 42.14 & \\
~Age: 50-65 & -1.64 &  & 16.82 &  & 171.49 &  & -89.45 &  & -2.85 &  & 94.37 & \\
~Sex: Male & \textbf{21.31} & \textbf{\textsuperscript{.}} & 26.12 &  & -6.29 &  & 3.17 &  & -16.17 &  & 28.13 & \\
~Race: Asian & \textbf{60.83} & \textbf{**} & 21.95 &  & -21.71 &  & 16.12 &  & -58.01 &  & 19.17 & \\
~Race: Black/Af. Am. & \textbf{84.02} & \textbf{**} & 79.44 &  & 8.09 &  & 36.66 &  & -37.26 &  & 170.95 & \\
~Race: Hisp./Latino &  \textbf{10.16} & \textbf{.} & -9.70 &  & 20.90 &  & -47.30 &  & 15.62 &  & -10.32 & \\
~Race: White & 50.23 &  & 5.15 &  & -52.88 &  & -6.08 &  & -65.60 &  & -69.17 & \\
~Education: Associate & 34.92 &  & 66.28 &  & 33.44 &  & \textbf{-57.75} & \textbf{*} & -44.99 &  & 31.89 & \\
~Education: Bachelor's & 48.75 &  & 40.93 &  & 67.25 &  & -62.80 &  & -41.46 &  & 52.66 & \\
~Education: HS/Diploma & 20.30 &  & 37.23 &  & 92.65 &  & \textbf{-100.01} & \textbf{*} & 1.70 &  & 51.88 & \\
\rowcollight \multicolumn{12}{l}{\textit{Psychological Traits}}\\
~BFI: Extraversion & 14.06 &  & -1.08 &  & -10.26 &  & -2.84 &  & -23.62 &  & -23.75 & \\
~BFI: Agreeableness & \textbf{-25.01} & \textbf{*} & 6.15 &  & 15.08 &  & \textbf{-28.74} & \textbf{**} & \textbf{49.48} & \textbf{*} & 16.97 & \\
~BFI: Conscientiousness & -3.15 &  & 15.72 &  & 9.54 &  & 5.86 &  & 35.08 &  & 63.05 & \\
~BFI: Neuroticism & \textbf{-27.55} & \textbf{***} & 4.77 &  & \textbf{-18.99} & \textbf{\textsuperscript{.}} & 4.88 &  & 43.91 & *** & 7.02 & \\
~BFI: Openness & 22.07 &  & -6.73 &  & 5.11 &  & -9.27 &  & -31.68 &  & -20.51 & \\
\hdashline
~SSRQ: Self-Regulation & 3.92 &  & -7.82 &  & 23.56 &  & 2.80 &  & 24.92 &  & 47.38 & \\
\hdashline
~BSCS: Self-Control & \textbf{-3.75} & \textbf{**} & -4.21 &  & \textbf{-4.15} & \textbf{*} & -0.31 &  & 1.38 &  & \textbf{-11.04} & \textbf{*}\\
\hdashline
~PANAS: P. Affect & 1.76 &  & 1.82 &  & 1.06 &  & \textbf{-2.79} & \textbf{**} & \textbf{-5.36} & \textbf{*} & -3.51 & \\
~PANAS: N. Affect &  -0.69 &  & -1.56 &  & 0.31 &  & -0.72 &  & 2.12 &  & -0.54 & \\
\rowcollight & \multicolumn{2}{l}{\n{R^2}=\textbf{0.34***}} &\multicolumn{2}{l}{\n{R^2}=\textbf{0.10*}} & \multicolumn{2}{l}{\n{R^2}=\textbf{0.10*}} & \multicolumn{2}{l}{\n{R^2}=\textbf{0.22***}} &\multicolumn{2}{l}{\n{R^2}=\textbf{0.21***}} & \multicolumn{2}{l}{\n{R^2}=\textbf{0.14*}} \\
    \bottomrule
    \end{tabular}
\label{table:eodgap_regression}
\Description[table]{This table summarizes linear mixed effects regression models examining how individual differences relate to the end-of-day estimation–actual (E–A) gap in app usage across six domains: Creativity, Entertainment, Productivity, Shopping, Social, and Aggregated. Independent variables include device type, number of study days, demographic characteristics (e.g., age, sex, race, education), and psychological traits (e.g., Big Five, self-regulation, self-control, affect scores). Each cell reports the regression coefficient and corresponding p-value, with bolded values indicating statistically significant associations. R-squared (R²) values are reported in the final row, showing the proportion of variance explained by each model.}
\end{table*}

\subsection{Linking Estimated--Actual Gaps with Daily Wellbeing}
We next examined how \ea{}s relate to daily self-reported wellbeing, including perceived 1) \textit{satisfaction}, 2) \textit{goal adherence}, 3) \textit{self-control}, 4) \textit{focus}, and 5) \textit{stress}. 
Similar to the above, we employed linear mixed-effects regression modeling, by building a separate model for each of the wellbeing measures as a dependent variable. 
For independent variables, we included individual-level factors (demographics, psychological traits, smartphone type, and estimation--actual gaps), and day-level controls (study day and weekend/weekday indicator).

\autoref{table:wellbeing_regression} reports the $\beta$-coefficients and statistical significance.  
Our results show that start-of-the-day (SoD) E--A gaps are significantly associated with all wellbeing outcomes except focus. 
Smaller gaps (i.e., more accurate self-estimation) are linked with higher satisfaction ($\beta$=-1.3E-3) and stronger self-control ($\beta$ = -1E-3), suggesting that perceiving one's digital use more accurately is associated with a sense of satisfaction and self-control.
However, smaller gaps are also associated with lower goal adherence ($\beta$=1.6E-3) and greater stress ($\beta$=-9E-4). 
\textbf{This observation highlights a key paradox that while accurate self-estimation may enhance satisfaction and control, it may simultaneously amplify the burden or stress associated with adhering to one's goals.}

Our qualitative findings further illustrate the paradoxical role of accurate estimation revealed in our regression models. 
For some participants, accurate awareness reinforced acceptance or neutrality, as one participant explained: \textit{``I know what my usage is generally\dots but I am not gonna change my behavior based on the settings digital wellbeing screen that I am seeing.'' (P21)} Still others noted how gaps sharpened their awareness without necessarily resulting in immediate change: \textit{``I had a larger social media usage than I anticipated. So that is something that I was a bit more aware of [that day].'' (P4)} 
 \edit{While smaller E--A gaps were associated with higher satisfaction and self-control, they were also linked with increased stress and lower goal adherence. Participants' narratives illuminate this paradox: several described feeling pressured to ``meet'' their predicted usage or experiencing guilt when their actual behaviors diverged from their forecasts, and others described the daily estimates as creating a perceived obligation to conform to their own expectations.}These observations 
highlight how accurate self-estimation could enhance satisfaction and self-control while also amplifying stress and goal conflict.

\begin{table*}[t]
\sffamily
\centering
\footnotesize
\caption{\textbf{Daily Wellbeing: }Linear mixed effects regression models revealing the relationship between individual differences and daily self-awareness of app usage with daily self-reports of feeling satisfied, difficulty, self-control, focused, and stressed; (\textbf{Bolded} values are for \textsuperscript{.} \n{p}\textless{}0.1, * \n{p}\textless{}0.05, ** \n{p}\textless0.01, *** \n{p}\textless{}0.001).}
\begin{tabular}{l r@{}l r@{}l r@{}l r@{}l r@{}l}
    \textbf{Dep. Variables} $\rightarrow{}$ & \multicolumn{2}{c}{\textbf{Satisfaction}} & \multicolumn{2}{c}{\textbf{Adherence}} & \multicolumn{2}{c}{\textbf{Self-control}} & \multicolumn{2}{c}{\textbf{Focus}} & \multicolumn{2}{c}{\textbf{Stress}}\\ 
    \toprule
    \textbf{Indep. Variables} $\downarrow{}$ & \textbf{Coeff.} & & \textbf{Coeff.} &  &\textbf{Coeff.} &  &\textbf{Coeff.} &  &\textbf{Coeff.} &  \\  
    \cmidrule(lr){1-1}\cmidrule(lr){2-3}\cmidrule(lr){4-5}\cmidrule(lr){6-7}\cmidrule(lr){8-9}\cmidrule(lr){10-11} 
~Intercept & \textbf{6.98} & \textbf{\textsuperscript{.}} & -4.76 &  & \textbf{6.86} & \textbf{*} & -0.41 &  & \textbf{7.67} &  \textbf{*}\\
~Smartphone: IOS & 0.21 &  & -1.42 &  & -0.36 &  & \textbf{-1.40} & \textbf{**} & 0.19 & \\
~Weekend & 0.14 &   & 0.11 & & -0.11 &  & 0.06 & & 0.11 & \\
~Num. Study Days & 0.01 &  & -0.01 &  & \textbf{0.03} & \textbf{**} & 0.01 &  & 0.01 &  \\
\rowcollight \multicolumn{11}{l}{\textit{Demographics}}\\
~Age: 25-35 & -0.15 &  & 0.71 &  & -0.95 &  & -0.14 &  & -0.35 & \\
~Age: 50-65 & 1.14 &  & 0.68 &  & -0.86 &  & \textbf{-2.68} & \textbf{\textsuperscript{.}} & 0.82 & \\
~Sex: Male & 0.13 &  & -0.24 &  & 0.21 &  & \textbf{-0.93} & \textbf{***} & \textbf{0.69} & \textbf{**}\\
~Race: Asian & \textbf{-1.14} & \textbf{*} & 0.83 &  & -0.23 &  & \textbf{0.87} & \textbf{\textsuperscript{.}} & -0.23 & \\
~Race: Black/Af. Am. & -0.67 &  & 0.31 &  & \textbf{-1.07} & . & 0.80 &  & 0.42 & \\
~Race: Hisp./Latino & 0.49 &  & 0.39 &  & 0.18 &  & \textbf{1.82} & \textbf{**} & 0.37 & \\
~Race: White & \textbf{-2.20} & \textbf{***} & \textbf{1.40} & \textbf{*} & -0.94 &  & \textbf{1.46} & \textbf{**} & \textbf{-1.10} & \textbf{*}\\
~Education: Associate & \textbf{-1.02} &\textbf{*} & \textbf{1.40} & \textbf{**} & \textbf{-1.20} & \textbf{*} & \textbf{1.19} & \textbf{*} & -0.64 & \\
~Education: Bachelor's & -0.65 &  & 0.94 &  & -1.08 &  & -0.62 &  & -0.73 & \\
~Education: HS/Diploma & -0.38 &  & 1.57 &  & \textbf{-1.84} & \textbf{*} & \textbf{2.06} & \textbf{**} & -1.14 & \\
\rowcollight \multicolumn{11}{l}{\textit{Psychological Traits}}\\
~BFI: Extraversion & \textbf{-0.60} & ** & \textbf{0.48} & \textbf{**} & -0.43 & ** & 0.14 &  & \textbf{-0.79} & 
\textbf{***}\\
~BFI: Agreeableness & 0.33 &  & 0.38 &  & -0.31 &  & 2.20 & *** & -0.21 & \\
~BFI: Conscientiousness & 0.28 &  & -0.33 &  & 0.19 &  & -0.17 &  & 0.22 & \\
~BFI: Neuroticism & -0.19 &  & 0.12 &  & -0.14 &  & -0.87 & *** & -4.9E-3& \\
~BFI: Openness & \textbf{-0.78} & \textbf{*} & \textbf{0.66} & \textbf{*} & -0.31 &  & 0.24 &  & \textbf{-0.73} & \textbf{*}\\
\hdashline
~SSRQ: Self-Regulation & 0.59 &  & -0.29 &  & 0.10 &  & -0.19 &  & 0.38 & \\
\hdashline
~BSCS: Self-Control & \textbf{-0.09} & \textbf{*} & 0.04 &  & 0.04 &  & \textbf{-0.13} & \textbf{***} & -0.02 & \\
~PANAS: P. Affect & 4.1E-3 &  & 0.03 &  & -0.04 &  & 0.08 & ** & -2.8E-3 &\\
~PANAS: N. Affect & \textbf{0.07} & \textbf{*} & -1E-3 &  & 0.01 &  & 0.07 & ** & 0.01 & \\
\rowcollight \multicolumn{11}{l}{\textit{App Usage Self-expectation (SoD Predictions)}}\\
~SoD E-A Gap: Creativity & -1.9E-3 &  & 1E-3 &  & 5E-4 &  & 6E-4 &  & -8E-4 & \\
~SoD E-A Gap: Entertainment & -1E-3 &  & 6E-4 &  & \textbf{-1.9E-3} & \textbf{*} & 1.1E-3 &  & \textbf{-2.4E-3} & \textbf{**}\\
~SoD E-A Gap: Productivity & \textbf{1.6E-3} & \textbf{*} & \textbf{-2E-3} & \textbf{**} & 2.9E-3 & *** & \textbf{-1.5E-3} & \textbf{\textsuperscript{.}} & \textbf{1.8E-3} & \textbf{*}\\
~SoD E-A Gap: Shopping & 1.5E-3 &  & -1.9E-3 &  & 5E-4 &  & -8E-4 &  & \textbf{3E-3} & \textbf{**}\\
~SoD E-A Gap: Social & -1.6E-3 & \textsuperscript{.} & \textbf{4E-3} & \textbf{***} & \textbf{-3E-3} & \textbf{***} & 9E-4 &  & \textbf{-2.5E-3} & \textbf{***}\\
~SoD E-A Gap: Aggregated & \textbf{-1.3E-3} & \textbf{***} & \textbf{1.6E-3} & \textbf{***} & \textbf{-1E-3} & \textbf{***} & 2E-4 &  & \textbf{-9E-4} & \textbf{***}\\
\rowcollight \multicolumn{11}{l}{\textit{EoD Estimation--Actual Gap in App Usage}}\\
~EoD E-A Gap: Creativity & 2.4E-3 &  & -8E-4 &  & -6E-4 &  & -8E-4 &  & 2E-4 & \\
~EoD E-A Gap: Entertainment & 6E-4 &  & 5E-4 &  & 7E-4 &  & -3E-4 &  & \textbf{2E-3} &  \textbf{\textsuperscript{.}}\\
~EoD E-A Gap: Productivity & \textbf{-2E-3} & \textbf{ \textsuperscript{.}} & \textbf{2.5E-3} & \textbf{**} & -1.1E-3 &  & -6E-4 &  & \textbf{-3E-4} & \textbf{***}\\
~EoD Reflect Diff: Shopping & -1E-3 &  & -2E-4 &  & -5E-4 &  & 2.5E-3 &  & 1.7E-5 & \\
~EoD Reflect Diff: Social & 4E-4 &  & \textbf{-2.7E-3} & \textbf{**} & \textbf{1.9E-3} & \textbf{*} & -8E-4 &  & 1.1E-3 & \\
~EoD Reflect Diff: Aggregated & 3E-4 &  & \textbf{-6E-4} & \textbf{*} & 4E-4 &  & 8.3E-5 &  & 3E-4 & \\
\rowcollight & \multicolumn{2}{l}{\n{R^2}=0.37***} &\multicolumn{2}{l}{\n{R^2}=0.44***} & \multicolumn{2}{l}{\n{R^2}=0.41***} & \multicolumn{2}{l}{\n{R^2}=0.36***} & \multicolumn{2}{l}{\n{R^2}=0.34***} \\
    \bottomrule
    \end{tabular}
\label{table:wellbeing_regression}
\Description[table]{This table summarizes linear mixed effects regression models examining the relationship between individual differences and daily self-awareness of app usage with five self-reported daily outcomes: satisfaction, adherence, self-control, focus, and stress. The independent variables include device type, number of study days, demographics (such as age, sex, race, and education), psychological traits (including Big Five personality traits, self-regulation, self-control, and PANAS affect scores), prediction gaps between actual and estimated app usage at the start and end of the day (across categories like creativity, entertainment, productivity, shopping, and social), and overall estimation gaps. Each cell reports the regression coefficient, with asterisks indicating significance levels. Bolded values represent statistically significant associations. A final row presents the R-squared (R²) values, representing the overall fit of each regression model to the respective outcome variable.}
\end{table*}

\section{RQ2: \ws{} for Digital and Holistic Wellbeing}

We now shift our attention towards our participants' understandings of digital wellbeing and its interplay with their holistic wellbeing. We focus on how \ws{} may have impacted their emotional wellbeing, and how an extended period of daily logging and self-reflection may have affected their initial conceptualizations of DW. In doing so, we address our \textbf{RQ2:} \textit{How does a self-reflection tool influence users' self-reported wellbeing and understanding of digital wellbeing?} 



\subsection{Psychological Effects of Using \ws{}}
\begin{figure}[t]
 \begin{subfigure}[b]{0.5\columnwidth}
    \centering
    \includegraphics[width=\columnwidth]{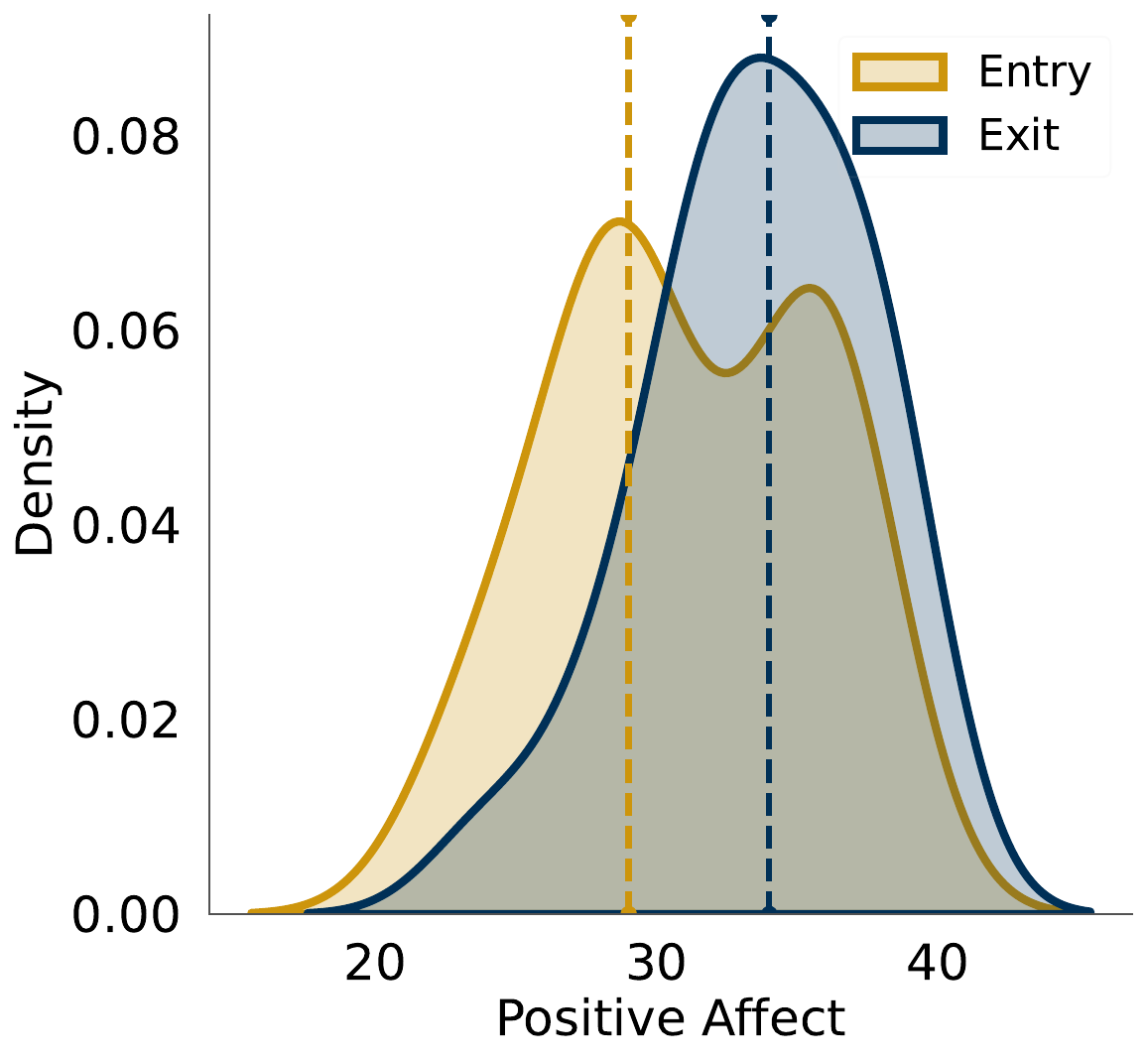}
    \caption{Pos. Affect}
    \label{fig:pos_affect}
    \end{subfigure}\hfill
\begin{subfigure}[b]{0.5\columnwidth}
    \centering
    \includegraphics[width=\columnwidth]{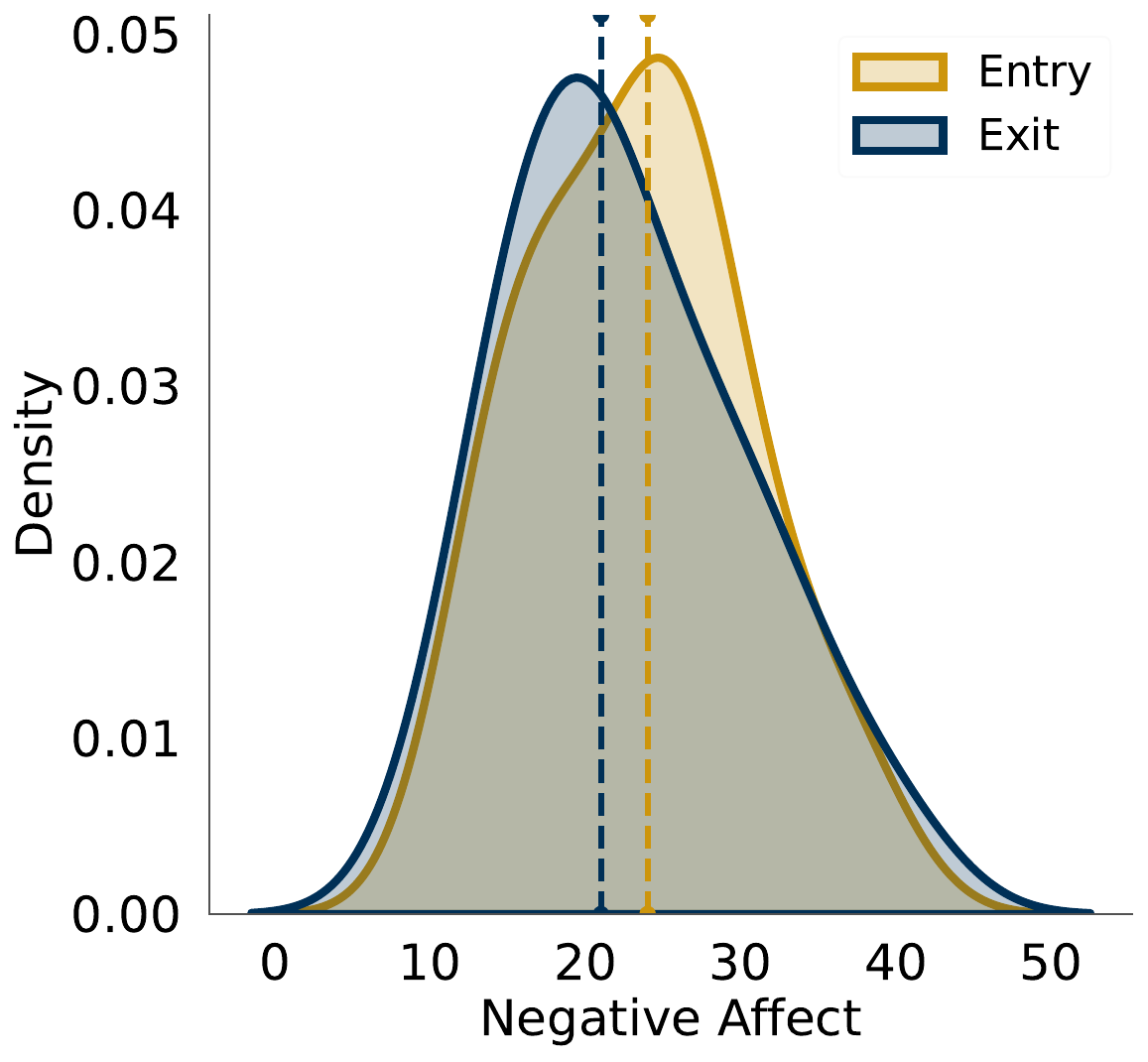}
    \caption{Neg. Affect}
    \label{fig:neg_affect}
    \end{subfigure}\hfill \begin{subfigure}[b]{0.5\columnwidth}
    \centering
    \includegraphics[width=\columnwidth]{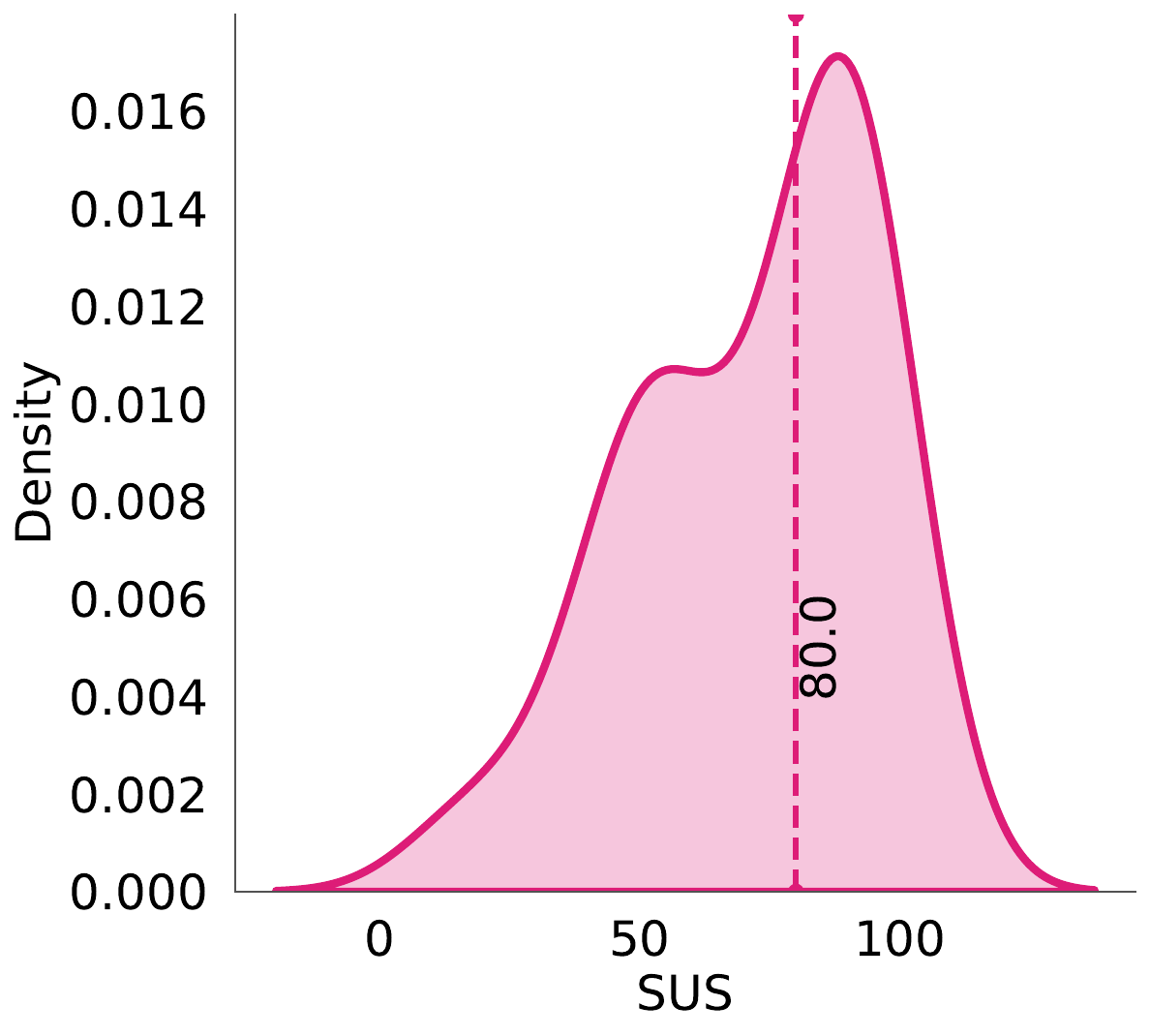}
    \caption{SUS}
    \label{fig:sus}
    \end{subfigure}\hfill
\begin{subfigure}[b]{0.5\columnwidth}
    \centering
    \includegraphics[width=\columnwidth]{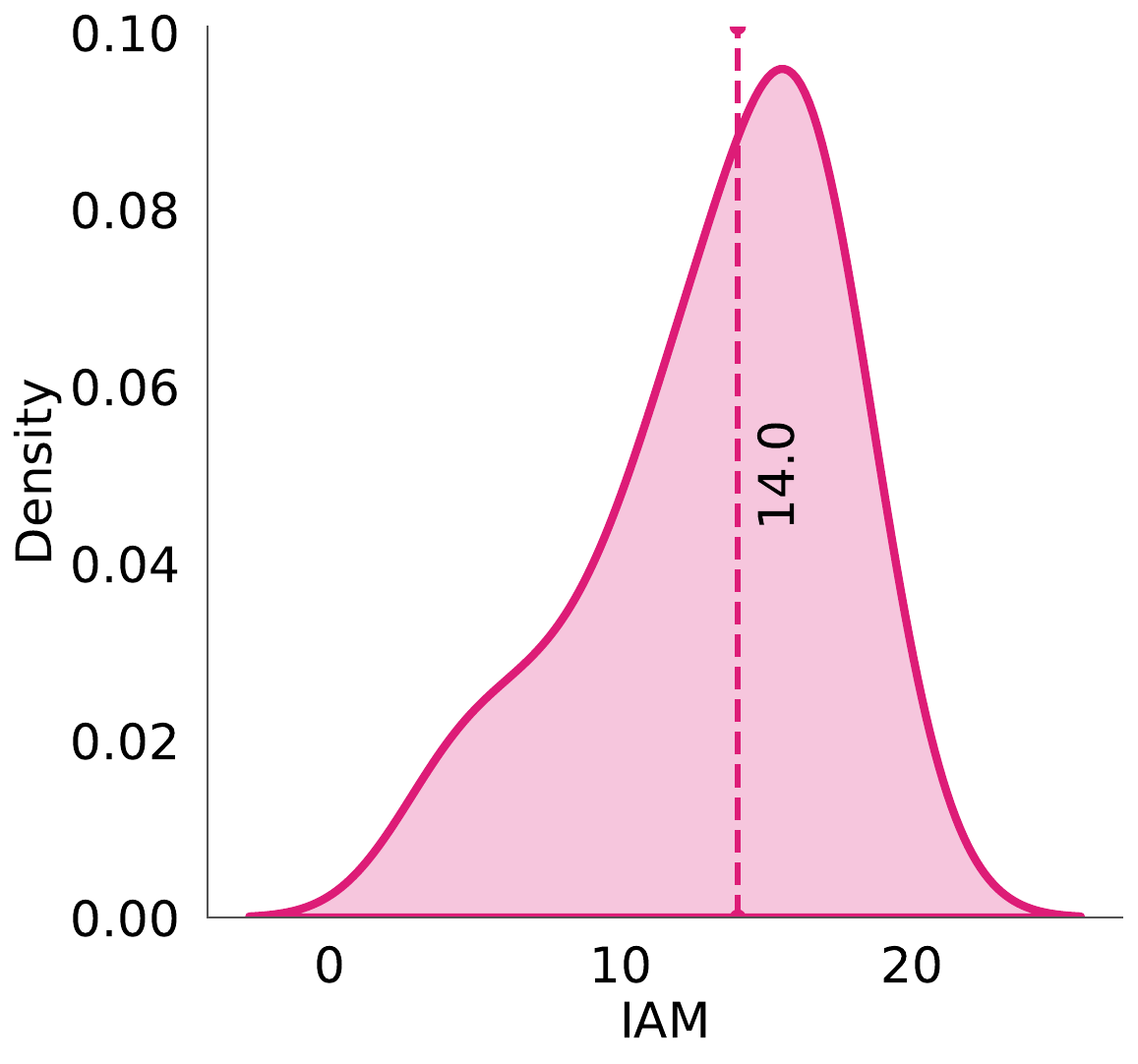}
    \caption{IAM}
    \label{fig:iam}
    \end{subfigure}\hfill
    \caption{Distribution plots based on the exit surveys (the dotted lines represent the median values of the corresponding color distribution).} 
    \label{fig:dist_plots}
    \Description[figure]{This figure shows the distribution of self-reported scores from exit surveys completed by participants. Plots (a) and (b) compare the distributions of Positive Affect and Negative Affect between the entry and exit timepoints, with gold representing entry and navy representing exit. Plots (c) and (d) show kernel density estimates for System Usability Scale (SUS) and Information Assessment Method (IAM) scores reported at the end of the study, with vertical dotted lines indicating the median values.  The corresponding dotted lines show the median of each distribution.}
\end{figure}

First, we examined within-person differences in participants' psychological assessments of self-regulation (SSRQ)~\cite{carey2004psychometric}, self-control (BSCS)~\cite{tangney2018high}, and emotional wellbeing or affect (PANAS-SF)~\cite{watson1988development,mackinnon1999short} before and after participation in the study (ref: \autoref{tab:psychological}). 
We find that the participants did not show significant changes in terms of self-regulation and self-control, which is not surprising given that both BSCS and SSRQ are considered to be immutable trait-based measures, capturing relatively stable dispositions that are unlikely to shift over the course of a short, two-week study. 
However, we see improvements in participants' emotional wellbeing after using \ws{}. 
In particular, participants showed a significant change in positive affect which increased by an average of 10.03\% with a large effect size (Cohen's $d$=0.63) and statistical significance ($t$=2.63, $p$<0.05) (ref:~\autoref{fig:pos_affect}). \edit{Our interview data indicates that these improvements might be driven less by behavioral change and more by shifts in emotional interpretation, as participants described accepting their existing habits and reflecting on perceived discrepancies. P7, for example, said:}
\begin{quote}
    \edit{``I feel like I'm more self-aware now, which is better because I didn't realize how much I was actually spending on some applications\dots and that really is helping me kind of structure what I do in a day.'' (P7)}
\end{quote}
Again, participants also showed a slight reduction in negative affect (-0.12\% on average, Cohen's $d$=-0.12), though this change was not statistically significant (ref:~\autoref{fig:neg_affect}). Taken together, these results suggest that \ws{} use may have contributed small but meaningful improvements in participants' day-to-day emotional wellbeing, primarily through improved positive affect.


\edit{This disconnect between improved emotional wellbeing and stable self-regulatory traits is consistent with psychological accounts of reflective meaning-making and self-appraisal. In their interviews, participants expressed acceptance of their habits and post-hoc justification of their E--A gaps. These mirror mechanisms of self-discrepancy reduction~\cite{Higgins_1989}, where reflection allows individuals to minimize distress by reconciling perceived gaps between their ``actual'' and ``ideal'' behaviors. Structured daily reflections may have also functioned as a form of meaning-making, allowing participants to create narratives around their technology use in ways that fostered emotional coherence and self-understanding~\cite{Mekler_Framework_2019}, even in the absence of measurable behavior change. Together, these mechanisms might help explain why \ws{} supported short-term improvements in positive affect while leaving trait self-control and self-regulation unchanged.} 





\begin{table}[t!]
\centering
\sffamily
\footnotesize
   \caption{Comparing psychological assessments at entry and exit of the study, with median values in entry and exit surveys, relative change ($\Delta$) \%, effect size (Cohen's $d$), and paired $t$-tests (\textbf{Bold} values are for * $p$<0.05, ** $p$<0.01, *** $p$<0.001). The length of horizontal bars represents the magnitude of $\Delta$ \%, where \textcolor{exitcol}{\textbf{BLUE}} bars indicate an \textcolor{exitcol}{increase} and \textcolor{entrycol}{\textbf{GOLDEN}} bars indicate an \textcolor{entrycol}{decrease} in the measure following study period.}   \label{tab:psychological}
\resizebox{\columnwidth}{!}{
\begin{tabular}{lrrr@{}c@{}lrr@{}l}
 \textbf{Measure} & \textbf{Entry} & \textbf{Exit} & &\textbf{$\Delta$ \%} & & \textbf{Cohen's $d$} & \multicolumn{2}{c}{\textbf{$t$-test}}\\
\toprule
BSCS: Self-control & 47.50 & 46.00 &~\entrybar{0.11}  & -0.11 &  & -0.05 & -0.30 & \\
SSRQ: Self-regulation & 3.55 & 3.50 & & 1.74 & ~\exitbar{1.74} & -0.04 & -0.21 & \\
PANAS: P. Affect & 29.00 & 34.00 & & 10.03 & ~\exitbar{10.03} & 0.63 & \textbf{2.63} & \textbf{**}\\
PANAS: N. Affect & 24.50 & 21.00 & \entrybar{0.12}~ &  -0.12 & & -0.12 & -0.53 &  \\
\bottomrule
\end{tabular}}
\Description[table]{This table presents median scores and standard deviations for four psychological assessments administered at the beginning (entry) and end (exit) of the study. The measures include self-control (BSCS), self-regulation (SSRQ), and emotional wellbeing captured by the PANAS-SF scales for Positive and Negative Affect. Each row corresponds to a distinct psychological measure. Columns display the median (Mdn.) at entry and exit, the percentage change ($\Delta$\%), effect size (Cohen's $d$), and the $t$-test statistic comparing entry and exit scores. Statistically significant $t$-test results are \textbf{bolded} and marked with asterisks to denote significance levels.}
\end{table}

\subsection{Evolving Definitions of Digital Wellbeing}
At the start of the study, our participants' definitions of DW were aligned with widely acknowledged definitions that centered on time management around digital technology usage and ensuring balance between their digital and physical lives. For example, P6 explained that DW was \textit{``mostly about time management''} and not spending \textit{``too much time on devices that it takes away from you doing other stuff,''} with P9 adding that it concerned, \textit{``the balance between your real life and your virtual life.''}

Following the 14-day intervention using \ws{}, we observed a notable change in how some participants defined DW for themselves. While core concepts of balance and time management remained, their understanding became more nuanced. For many, the definition evolved to include a greater emphasis on 
the emotional impact of digital interactions. P5 noted that, for her, DW meant \textit{``being intentional and more more mindful of usage of digital stuff, so that you feel like you're benefiting from it.''} Having initially described DW through a narrow focus on avoiding negative emotions, P25 reflected on his renewed understanding of DW as:
\begin{quote}
\small
``I think that DW involves intentionally using my phone for purposes that are useful, and in the case of entertainment, ensuring that content and social interactions are positive and don't incite negative emotions. This includes avoiding mindless content, such as short-form videos on platforms like YouTube and Snapchat, and steering clear of platforms, like Twitter, that previously provoked feelings of anger or frustration.'' (P25)
\end{quote}

Diverging from these more traditional definitions, some participants expanded the scope of what they considered DW to include their digital footprints in our increasingly digital worlds. They expressed that DW was not only about their personal balance or time management of technology usage but also cognizance of the ripple effects of their digital presence beyond themselves to include their portrayals of themselves to the world. P4 summarized this as:
\begin{quote}
    \small
    ``I believe digital wellbeing means that you have control over the information that you're sharing\dots and this can be with AI systems,\dots search engines,\dots corporations like Google. Having control and a healthy amount of awareness about the information and the digital footprint that you're putting out into the world\dots That is digital wellbeing for me.'' (P4)
\end{quote}
This shift from a focus on personal device usage to a broader concern for one's digital presence--including information shared with corporations and AI--underscores why self-awareness is a critical and necessary component of the DW narrative. It suggests that true DW requires not only conscious management of one's time and emotions but also an awareness of the lasting impact of one's digital actions on oneself. 

\subsection{Intersections of Digital and Mental Wellbeing}
We identified several patterns of technology use that participants described as having a direct impact on their mental wellbeing. In particular, we noted behaviors that were often habitual, performed with little conscious thought, that participants expressed as leading to negative feelings associated with loss of control over their time as well as potentially skewed screen time calculations. 
The first, and most common, pattern was the use of devices at the bookends of the day--first thing in the morning and last things at night: 
\begin{quote}
\small
``I'm not sure how it started, but scrolling for 30 minutes before bed has become an unconscious habit that's now a regular part of my routine.'' (P16)
\end{quote}

We note that this expressed need for background entertainment 
may create a state of perpetual connectivity, where ``downtime'' is often filled with habitual phone use, a behavior that many participants acknowledged made them feel out of control and negatively impacted wellbeing.
For example, P9 described using technology to disconnect from the world and avoid negative feelings, stating: 
\begin{quote}
\small
``If I'm just sad or if I'm not feeling good, I probably run up to YouTube or Netflix to turn my brain off, so I don't have to think about it. I can get lost in that world. I don't have to think about the sadness or what I'm mad about.'' (P9)
\end{quote}
It is ostensible that such digital device usage habits functioning as coping mechanisms may provide short-term benefits. However, our data surfaced that they do contribute to a long-term loss of control. Our participants described similar behaviors in social settings, where they would pick up their phones during lulls in conversation with friends. These actions highlight a pervasive need to fill every moment with stimulation, which might have cognitive burdens and overloads that are not captured by traditional screen time metrics. These habits may have incommensurate effects on mental health compared to the relatively smaller effect on their screen time. 



\section{RQ3: \ws{} for Digital Self-Awareness}

Here, we present our analysis of how \ws{}, designed as a self-reflection tool, enabled our participants to engage in more mindful smartphone usage behaviors towards great self-awareness. In doing so, we address our \textbf{RQ3:} \textit{How do users perceive the usefulness of a self-reflection tool, and how does it support digital self-awareness?} 



\subsection{Perceived Usefulness of \ws{}}

\subsubsection{Usability and Appropriateness of \ws{}} During the exit survey, participants completed the System Usability Scale (SUS; 0-100)~\cite{bangor2008empirical} and the Intervention Appropriateness Measure (IAM; 4-20)~\cite{weiner2017psychometric} to evaluate their experience in terms of usability and appropriateness with \ws{}.
The distribution plots are shown in \autoref{fig:sus} (for SUS) and \autoref{fig:iam} (for IAM).
The median SUS score was 80 (mean=71.91), which, consistent with prior work, is considered above average and acceptable usability~\cite{bangor2008empirical}. 
The median IAM score was 14 (mean=13.24), indicating that participants perceived the intervention as \textit{moderately appropriate}~\cite{weiner2017psychometric}. 
This suggests the intervention is viewed as a reasonable fit with potential value, though some issues with relevance or contextual alignment may warrant further exploration.

However, considering that this was not a high-fidelity prototype or a system---participants also expressed concerns with the technical issues and glitches in \ws{}, which may have affected the SUS score. Such feedback included:
\begin{quote}
\small
    ``I wasn't the biggest fan of the app\dots it was kind of hard to properly navigate, and it was kind of glitchy.'' (P24)    
\end{quote}
At the same time, participants also noted the usefulness of \ws{} in their daily self-reflections as \textit{``this study has made me realize that I don't have a great grasp on [my smartphone usage habits] as I thought I did, but I'm hoping to\dots find that balance.'' (P7)} 



\subsubsection{A Two-dimensional Lens on Enthusiasm about Future Use}

It would have been interesting to examine how different individual-level factors shaped participants' perceptions of \ws{} using robust statistical analyses. However, given the small sample size of those who completed the study ($N$=20), participant-level quantitative tests were not statistically meaningful. 
Instead, we conducted a descriptive post-hoc analysis focusing on two factors central to digital wellbeing in our findings: self-control (which emerged as significant in our regression models; \autoref{table:sodgap_regression}, \autoref{table:eodgap_regression}, \autoref{table:wellbeing_regression}) and perceived usability (measured via SUS).

\begin{figure*}[t!]
    \centering
    \includegraphics[width=1.95\columnwidth]{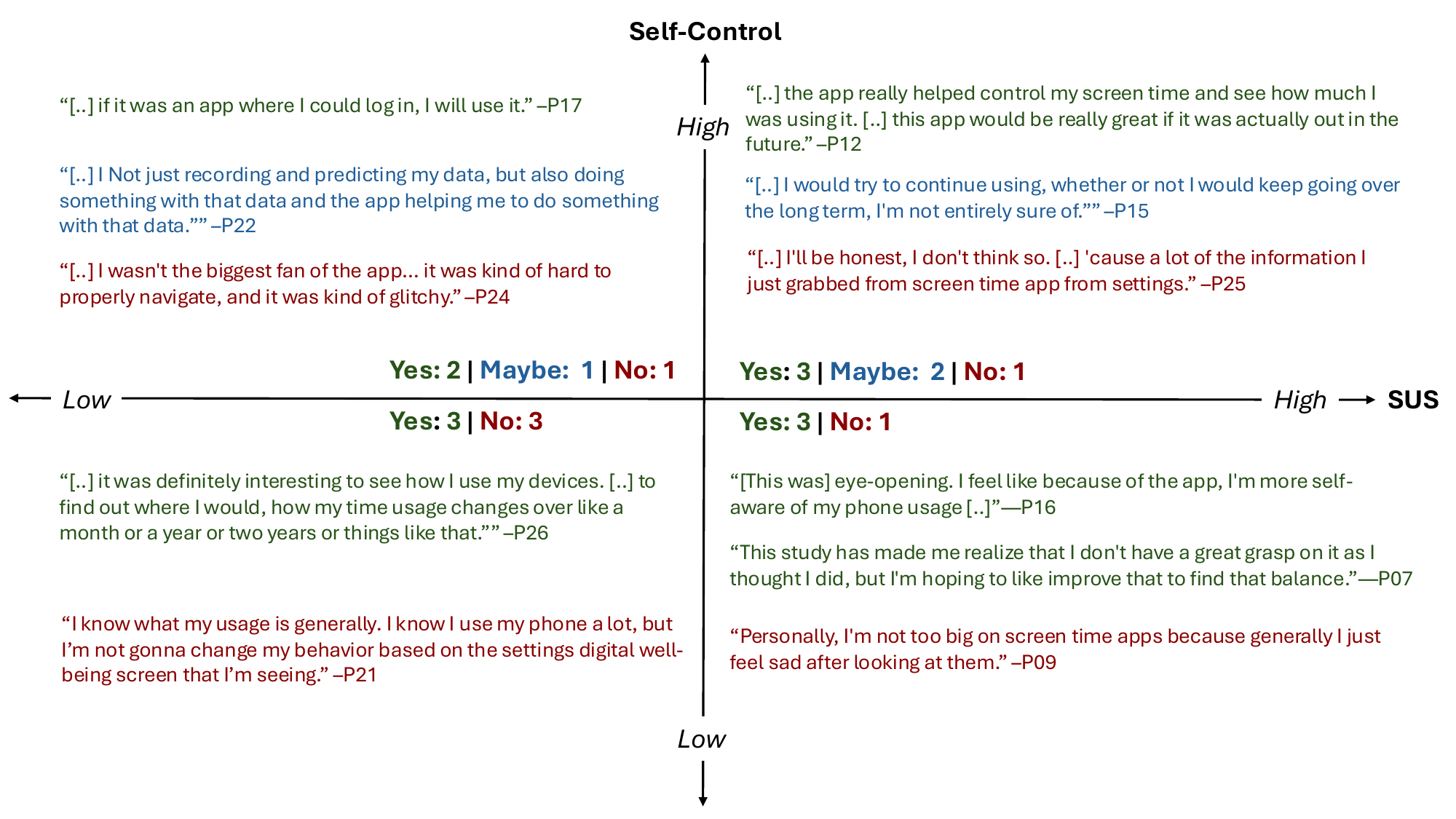}
    \caption{A two-dimensional representation showing distribution of participants by self-control (above vs. below median) and SUS usability scores (above vs. below median). The figure also shows example responses to the question ``Would you like to use this tool in the future?'' (\textcolor{dgreen}{\textbf{Yes}}/\textcolor{darkred}{\textbf{No}}/\textcolor{dblue}{\textbf{Maybe}}), along with example responses in each quadrant.}
    \label{fig:2d_plot}
    \Description[figure]{This figure displays a quadrant layout organizing participant quotes based on two axes: self-control (vertical axis, increasing from bottom to top) and perceived usability of the WellScreen app as measured by SUS scores (horizontal axis, increasing from left to right). Each quote represents a participant's response about whether they would continue using the app. Responses are color-coded: green for “Yes,” blue for “Maybe,” and red for “No.”}
\end{figure*}

As shown in~\autoref{fig:2d_plot}, we mapped participants onto a two-dimensional space split by median self-control (vertical axis) and median SUS (horizontal axis).
\edit{We used the median as the decision boundary for interpreting SUS and self-control scores because it offers a conservative and robust measure of central tendency that is less sensitive to outliers and skewed distributions, providing an interpretable threshold for comparing relative levels of these constructs within our sample.}
Within each quadrant, we overlaid their responses to whether they would like to continue using \ws{} in the future (Yes/No/Maybe), along with illustrative quotes from exit interviews. We describe our observations below:

\para{Quadrant 1 (High self-control X High SUS):} Most participants expressed willingness to continue using (3 Yes, 2 Maybe), though P25 was concerned about putting extra effort for using \ws{}, ``It didn't do enough for me---it was me putting in all the effort.''

\para{Quadrant 2 (High self-control X Low SUS):} Responses were mixed (2 Yes, 1 Maybe, 1 No). P22 in this quadrant desired more advanced functionality, ``Not just recording and predicting my data, but also doing something with it.''

\para{Quadrant 3 (Low self-control X Low SUS):} Opinions were evenly split (3 Yes, 3 No). While P26 valued awareness, ``It was interesting to see how my usage changes,'' P9 described negative or demotivating experiences, ``I just feel sad after looking at them.''

\para{Quadrant 4 (Low self-control X High SUS):} 
This quadrant showed the strongest enthusiasm (3 Yes, 1 No). P16 explicitly emphasized about ``self-awareness,'' while P9 reported feeling sad after looking at screen times.

This descriptive mapping underscores how future-use intentions or enthusiasm are shaped by the interplay of personal traits and perceived usability. This highlights that enthusiasm for \ws{} emerges when need (low self-control) aligns with usability (high SUS). Where either factor is missing, enthusiasm falters---either because \ws{} feels unnecessary (high self-control) or because the interface undermines potential value (low SUS).

\subsection{Unpacking Self-Awareness through Self-Reporting and Reflection}


\subsubsection{Mechanics of Self-Reporting and Self-Reflection: A Focus on Screen Time}

Arguably, screen time---the duration of time that a device's screen is on in a day---is a sufficient proxy for the time a user spent on their smartphones, leading to its emergence as the primary dimension along which digital wellbeing is measured. Our data helped us uncover several user patterns that introduced additional nuance to this conversation, allowing us to discuss the processes of encouraging self-reporting and reflection with screen time data towards better self-awareness.

First, several participants reported their smartphone usage behaviors that would have rendered their screen time an incomplete measure of the time they spent actually interacting with their phones. Multiple participants reported falling \textit{``asleep with TikTok open, because I was not awake from 4am to 9:30am.''} Along similar lines, many participants noted desiring some form of media or entertainment as ``white noise'' as they went about their daily activities. To them, this created a sense of being constantly \textit{on}, as their smartphones were always running in the background while they were doing other things. P10 described this as:
\begin{quote}
\small
``I find that having media playing in the background enhances my focus, particularly when I'm engaged in routine tasks such as handling emails. It makes these activities feel less monotonous and more enjoyable. I've also found this approach to be effective when I'm studying'' (P10)
\end{quote}

However, this behavior presents a significant challenge for data logging and awareness because the available data could often be unreliable. For example, participants who used YouTube for background music had to keep screens on, leading to inflated screen time numbers. In contrast, participants who used Spotify for the same purpose could play music with the screen off, resulting in a more accurate reflection of device usage but not overall engagement. As P27 noted, this behavior skews the screen time metric: 
\begin{quote}
\small
``I didn't have many goals yesterday, more relaxing. I did listen to Spotify for one and a half hours, but screen time doesn't show that since the screen is not on while listening.'' (P27)
\end{quote}

These insights reinforce a key shortcoming of employing screen time as the primary metric for assessing an individual's digital wellbeing. With the numbers not affording explanations or accommodations for annotations and added nuance, our participants were hindered in their ability to accurately reflect on their smartphone usage behaviors.


\subsubsection{Awareness of Intentionality: Value Judgment of Digital Behaviors}

We examine how our interviews surfaced participants' intentionality in their technology use and how it stood in contradiction with their emotions elicited by screen time metrics. 

Our participants consistently distinguished between ``good'' or ``meaningful'' use and ``bad'' or ``mindless'' use, often regardless of the total screen time logged by their devices. These value judgments were tied to activities they considered productive, socially valuable, or personally enriching. For instance, P4 categorized their time spent on industry news as a form of productivity, stating, 
\begin{quote}
\small
``I usually spend a significant amount of time catching up on news and catching up on stuff that's happening in my industry that I'm hoping to work in. I personally categorize that towards productivity and like work.'' (P4)
\end{quote}
Conversely, mindless or unintentional use was a source of frustration. P22 explained that he \textit{``just automatically, for some reason, find myself opening Messenger to check if any of my friends have messaged me, or I automatically open Instagram,''} expressing dissatisfaction at his daily digital habits and his perceived lack of agency in changing such behaviors.

At the same time, a single platform could serve entirely different functions for different users or for any single user under different circumstances. Our data surfaced examples including how, Instagram could be used for ``doomscrolling'' or to mindlessly pass time, but it could also be used to plan a social event with friends, which holds a vastly different personal value. However, screen time data adds all of these minutes up as a single, undifferentiated metric. 
Indeed, several participants reported that their data elicited feelings of guilt and disappointment, even for what they considered meaningful activities like connecting with loved ones or preparing for school.

As a result of this dichotomy, participants often felt they needed to defend their screen time numbers while reporting their daily estimations on \ws{}. The data, while intended to promote self-awareness, frequently created a sense of guilt or a need for justification. We noted in the daily reflections that several participants expressed disappointment around their numbers---especially when they strayed greatly from their estimated figures. In cases where app categories, as described above, combined ``good use'' (staying in touch with a loved one) and ``bad use'' (mindless scrolling on social media), participants may have felt the need to set the record straight for themselves in their self-reflections. P18 captured this best, reporting:
\begin{quote}
\small
    ``OKAY IN MY DEFENSE, FIVE OF RHOSE (sic) HOURS ARE SNAPCHAT AND ONLY THREE OF THEM ARE \\ INSTAGRAM.'' (P18)
\end{quote} 

Lastly, we recognize that the observer effect (e.g.,~\cite{kazdin1982observer, saha2024observer}) might have influenced some of our participants' smartphone usage behaviors and self-reflections during their participation in our study. We found that some participants \textit{``unconsciously used [their smartphones] less, hoping that I wouldn’t hit those predictions,''} whereas others expressed that they thought twice before picking up their phone each time because it would add to their screen time. On the other hand, P17 chose to consider their estimations as daily goals, explaining:
\begin{quote}
    \small
    ``every time I made a prediction… I would make conscious choices to be like, OK, I spent quite a bit of time on social media yesterday, so maybe I should predict less or hopefully I use less.'' (P17)
\end{quote}

In sum, participants seemingly examined their data and then re-evaluated their actions from the previous days, often finding a way to reconcile a high number with an activity they deemed valuable. This captures a crucial finding that the self-reporting process of \ws{} didn't simply elicit a number to calculate an estimated--actual gap, it initiated a process of judgment and justification. To our participants, \ws{} served as a mirror, forcing them to confront a number they often saw as negative and then explain why, in their eyes, it was not.


\section{Discussion}\label{sec:discussion}


Our study examined \ws{} as a lightweight intervention to support digital self-awareness by prompting participants to estimate their daily smartphone use and reflect on the discrepancies between estimated and actual use (E--A gap). 
We scoped our study to college students, who participated in the study over two weeks. 
Our findings revealed interesting patterns in the E--A gap across app categories, nuanced associations between E--A gaps and wellbeing, as well as participants' reflections and explanations behind these patterns.
In this section, we discuss how our findings contribute to self-awareness and digital wellbeing literature, highlighting theoretical, practical, and design implications for future tools.

\subsection{Self-Reflection for Self-Awareness: Beyond Screen Time}





\subsubsection{The Paradox of Smaller Estimated--Actual Gap}
In theory, smaller estimated--actual (E--A) gaps indicate closer alignment between individuals' goals and routines in digital use, and should therefore be uniformly beneficial.  
However, our findings highlight a more paradoxical role of the E--A gap. 
While smaller gaps did coincide with higher satisfaction and self-control, they also coincided with greater stress as well as difficulty in goal adherence. 
This paradox aligns with prior work on ``good enough'' goal achievement, which argues that success should allow for margins rather than demand perfection~\cite{jung2021good}. 
Similarly, research on starting with goals suggests that discrepancies matter less as absolute measures and more in relation to what people hope to achieve~\cite{ekhtiar2023goals}.
In our case, the \ws{} intervention may have implicitly encouraged participants to strive for minimal gaps. Such striving plausibly fostered greater self-control, but also introduced additional stress. Alternatively, our findings point to rare instances where our participants may have intentionally chosen not to log their estimates and actual usage on days where they anticipated large E--A gaps, forestalling the stress altogether.


That said, the causal direction remains ambiguous---it could also be plausible that on days when individuals are already stressed or burdened with heavier workload, they may have exhibited better digital self-awareness, and as a result, achieved smaller E--A gaps.
This ambiguity underscores that smaller gaps should not be interpreted solely as outcomes of improved regulation, but also as potential byproducts of contextual pressures. Understanding this dynamic requires moving beyond static associations to disentangle how self-awareness practices interact with fluctuating states like stress, workload, and goal pursuit.

\subsubsection{From Screen Time to Meaningful Use}
Reflecting back on our motivation in \autoref{section:intro}, much of the digital wellbeing literature has centered on restriction---using less technology, spending fewer hours online, or controlling impulses through timers and goal-setting~\cite{ko2015nugu,zhou2021time,lyngs2019self,nguyen2021managing}. While these approaches can be beneficial, we argued that ``screen time'' is neither a sufficient nor always a practical basis for wellbeing. Our findings reinforce this concern. 
Participants both under- and over-estimated their use depending on app category, revealing the blurry boundaries between productivity, entertainment, and social interactions. Such discrepancies highlight the limits of treating screen time as a direct proxy for wellbeing.
This supports calls to reframe digital wellbeing not around screen time reduction, but around fostering awareness of how technologies support or hinder valued goals.

Along similar lines, our results point to self-awareness as a more meaningful pathway toward digital wellbeing. By prompting participants to estimate and reflect on their own digital use, \ws{} encouraged them to consider how their routines aligned or misaligned with personal goals on digital use. 
Some participants explicitly described the benefits from this practice, whereas some noted its drawbacks like experiencing worse moods on learning their screen times. 
Such reactions suggest that self-awareness should not be narrowly equated with screen time metrics. Instead, it could be broadened to include reflections on tasks, goals, and meaningful activities, allowing individuals to situate their digital practices within the wider contexts of daily life. This perspective resonates with broader shifts in digital wellbeing research, which emphasize mindful and value-driven engagement rather than blanket reduction~\cite{lukoff2018makes,thatcher2018mindfulness,saha2025mental}.



\subsubsection{Unequal Benefits and Risks of Misapplication}
Our analyses suggest that \ws{} is most beneficial for individuals with certain psychological traits. For example, individuals with lower self-control could be benefited to have lower E--A gap and higher positive wellbeing. 
However, those with higher self-control may already be good with managing their digital use better and may rather be burdened with using an intervention such as \ws{} on a daily basis.
This heterogeneity resonates with prior findings that self-tracking tools may benefit some individuals more than others~\cite{Epstein_Beyond_2016, bhat2020sociocultural}.
Further, although \ws{} achieved above-average usability, participants rated it as only moderately appropriate. This indicates that such tools could be functional yet still misaligned with lived contexts. Work schedules, productivity--leisure tradeoffs, and relational uses of technology may complicate whether interventions feel personally relevant to users.

These differences raise important questions for practitioners and policymakers: who benefits from wellbeing tools and who might be left behind? 
When technologies like \ws{} are rolled out in institutional settings (e.g., schools or workplaces)---even with the best of intentions---their impact may be uneven. Some individuals may experience genuine reflection and growth, while others could find the tool redundant, stigmatizing, or even punitive---especially if it is used for performance assessment. 

To avoid these pitfalls, digital wellbeing interventions must be designed and deployed as flexible, inclusive systems that accommodate diverse user needs and goal orientations.
A forward-looking path could involve participatory design approaches—engaging stakeholders directly in co-defining the purpose, goals, and evaluation metrics for wellbeing tools in organizational contexts, ensuring they foster empowerment rather than reinforcing inequities~\cite{kawakami2023wellbeing}.





\subsection{Design Implications}

\subsubsection{Balancing Self-Awareness and Self-Compassion}
As we look ahead to design new technologies to support digital self-awareness, interventions should position E--A gaps as opportunities for reflection rather than as ``errors'' to be corrected. 
If gaps are framed as failures, users risk experiencing guilt or pressure to achieve ``perfect'' alignment, which can undermine wellbeing. 
In contrast, framing discrepancies as natural and informative can support more sustainable engagement, encouraging users to learn about their patterns without judgment. 
For example, instead of interpreting higher-than-estimated productivity use as a failure, users might be encouraged to record a quick voice note acknowledging that they nevertheless accomplished important tasks (e.g., \textit{``I finished x, y, and z even though I spent more time online''}). 
Similarly, rather than treating more entertainment use as wasted time, tools could allow users to treat time as a budget distributed across days. 
A lower-than-estimated entertainment day could be balanced by greater entertainment use later in the week, reframing engagement as a flexible plan rather than a daily pass/fail test.


Designing such features could help normalize fluctuations in use, reduce guilt associated with ``off'' days, and foster a more compassionate, long-term perspective on digital routines. 
This balance is critical for avoiding the pitfalls of perfectionistic self-tracking and instead promoting compassionate awareness practices that people can maintain over time, supporting reflective practices that encourage users to celebrate accomplishments, adjust expectations, and maintain healthier relationships with technology.

\subsubsection{Embedding Prediction–Reflection Workflows}
The manual act of estimating digital use and later reconciling those estimates with outcomes encouraged reflections and emotional wellbeing. 
Importantly, the value was not only in the accuracy of predictions, but in the act of pausing, anticipating, and later evaluating. Such moments of self-estimation can function as micro-reflections that bring digital behaviors into conscious awareness.
Future tools can build on this by embedding lightweight estimation--reflection workflows into everyday routines. For instance, a morning prompt might present a projected estimate based on past habits (e.g., \textit{``We expect you may use your phone $\sim$3 hours today---does this match your plan?''}), inviting the user to confirm or adjust expectations. At the end of the day, a reflective prompt could encourage users to compare actual use with their morning estimate and annotate the reasons for discrepancies.

These workflows can be expanded to go beyond raw time. Users could be asked to predict the balance of activities (e.g., productivity, social, entertainment) or their intended goals (e.g., `\textit{`I want to spend more time connecting with friends than scrolling aimlessly''}). EoD reflections could then help them evaluate whether their digital use supported those intentions. Such workflows transform E--A awareness from a numerical comparison into a habit of intentionality, helping users contextualize their digital routines in relation to personal priorities. By weaving prediction and reflection into daily life, interventions can scaffold ongoing awareness without demanding perfection. Over time, these interventions may support greater intentionality, reinforce positive routines, and enable individuals to treat their digital practices as flexible and evolving rather than rigidly judged against single-day targets.


\subsubsection{Re-envisioning Digital Self-Awareness Tools for Holistic Wellbeing}
Our findings highlight the importance of designing digital wellbeing tools that move beyond rigid categorizations and narrow behavioral targets. Participants' estimation–actual (E--A) gaps revealed how apps often serve overlapping roles---a platform used for downtime may simultaneously provide social connection, while productivity apps may feel both enabling and burdensome. Interventions that label such use as simply ``good'' or ``bad'' risk misrepresenting lived experiences. Instead, tools should be sensitive to context and purpose, recognizing that the same digital activity can alternately support or strain wellbeing depending on circumstances.
Further, such interventions should not focus narrowly on reducing screen time or enforcing control, but rather on fostering everyday experiences of balance, satisfaction, and mindful engagement.

\subsubsection{Journaling and Summarization for Contextual Reflection}
While daily estimations and reflections are valuable, they can feel fragmented if not connected into broader insights or given sufficient context. A key pathway for deeper digital self-awareness is journaling~\cite{kim2024diarymate}. Even short, in-the-moment micro-entries—quick voice notes, brief text logs, or calendar-integrated prompts—can capture the why behind digital behaviors---what users were doing, how it made them feel, and how those experiences related to their goals. Such reflections provide essential context that raw usage data alone cannot convey~\cite{petelka2020being}.
These annotations can be paired with summaries over time that help individuals see how everyday fluctuations fit into evolving routines. For example, a weekly digest might highlight patterns such as: `\textit{`On days when you recorded task completion, your satisfaction scores were higher''} or \textit{``Entertainment use increased mid-week, which you linked to relaxation after deadlines.''} By synthesizing behaviors and reflections, such summaries help reframe digital practices as part of broader wellbeing trajectories rather than isolated daily outcomes.

\subsubsection{AI for Co-Interpreting and Playful Interrogation of Data}
The growing advances in generative AI and large language models (LLMs) present opportunities to make reflection more interactive, engaging, and sustainable~\cite{chopra2025engagements}. 
Rather than treating usage metrics as static outputs, AI can transform self-tracking into a dialogue where users actively interrogate, reinterpret, and even play with their own data.

One potential direction is to enable individuals to \textbf{question their data in natural language} (e.g., \textit{``Did I actually relax more on days when I spent extra time on entertainment?''} or \textit{``How did my mood shift when I balanced productivity and social apps?''}). 
Such exchanges position the system not as a tool or a judgmental authority but as a conversational assistant that helps surface patterns and test personal intuitions.

AI can also take a more \textbf{proactive role in co-interpreting data} by posing reflective or exploratory questions back to users. For instance---\textit{``I noticed your entertainment time was lower than usual—was that intentional or situational?''} or \textit{``Last week you logged feeling satisfied on high-productivity days---do you think that will hold true this week?''} These prompts encourage deeper meaning-making and help users link behaviors with contexts and values.

Beyond serious reflection, AI can infuse \textbf{playful activities} that make engagement with digital wellbeing less burdensome and more motivating. Technologies can invite users to make predictions about their future routines and then playfully ``fact-check'' those predictions, or present lighthearted trivia questions grounded in their own data (e.g., \textit{``Guess which day you spent the most time on social apps?''} or \textit{``Can you remember which activity boosted your satisfaction most this week?''}). This transforms data into a source of curiosity and self-discovery rather than guilt.
AI can also support gamified challenges that reframe progress as exploration rather than discipline (e.g., \textit{``This week you managed to increase your downtime while keeping stress low---want to see if you can repeat it three days in a row?''}). 
By blending interpretation with play, these activities normalize fluctuations in digital use while reinforcing flexible, compassionate self-awareness practices.

Taken together, these designs position AI not merely as a summarizer of data, but as a co-interpreter, conversational partner, and playful guide---one that helps users interrogate patterns, surface insights, and stay motivated to reflect. Rather than centering on accuracy or restriction, such designs emphasize flexible, compassionate, and personally meaningful engagement. These features point toward a broader design agenda for digital wellbeing---interventions that situate technology use in context, adapt to shifting purposes and meanings, and foster immediate affective benefits while supporting longer-term growth.

\subsection{Limitations and Future Directions}




Although our study provides important insights into how lightweight self-reflective interventions can support digital self-awareness, it has limitations, which also suggest interesting future directions.
\edit{This study involved a short, preliminary deployment to generate formative insights and did not include a control or comparison condition. While we acknowledge that this limits causal claims about the tool's effects, the study was not intended as a controlled experiment. Instead, our goal was to examine early usability, acceptability, and feasibility during a short (two-week) real-world deployment. A more rigorous controlled evaluation---such as 
a randomized field trial with a more functional prototype---will be an important next step to systematically assess the tool's impact on behavioral or psychosocial outcomes. Such a study design will also help disentangle the effects of confounding and latent external factors, such as self-monitoring~\cite{snyder1979self} and novelty effects~\cite{elston2021novelty}.}
Our study was limited to U.S. college students (25 participants, 20 completing). This modest sample yielded valuable insights but constrains generalizability. Different populations use and interpret technologies in varied ways, and future work should scale to larger and more diverse groups---including adolescents, working professionals, and older adults---to examine whether reflective scaffolds resonate across and between these groups and what design adaptations may be needed.


Further, our intervention was short-term and exploratory. 
A two-week deployment allowed us to capture meaningful shifts in emotional wellbeing, but longer-term longitudinal studies are needed to understand whether reflective practices lead to durable changes in digital habits and wellbeing. 
We also note the potential influence of novelty and observer effects~\cite{shin2019beyond,saha2024observer}---participants may have been immediately influenced by their awareness of using a new tool and by being observed, with social desirability potentially shaping how they monitored their smartphone use during the study. 
In addition, as \ws{} was deployed as a lightweight probe rather than a polished system, technical glitches and usability issues may have shaped participants' experiences and evaluations. \edit{Using platform-default app categories introduces potential mismatches in how our participants used certain applications and how they were characterized (\textit{e.g.,} Instagram is categorized as Social Media while, for a content creator, it might in fact be a Productivity tool). We attempted to ameliorate this issue by capturing any such potential mismatches through our daily reflections and interviews.} Iterating on design fidelity and integrating reflection more seamlessly into daily routines could enhance engagement and reduce friction. 

Finally, our study inspires a broader and deeper consideration of how digital wellbeing is interpreted and measured. 
While our intervention's scope was limited to screen time as a proxy for digital use, which---while useful as a baseline---overlooks important nuances such as the quality of digital content, the emotional tone of interactions, and the social or contextual settings of technology use. 
Our findings also underscore the need to account for personal and contextual factors that shape how individuals engage with technologies. 
Future interventions that incorporate these dimensions can move beyond simplistic measures of time spent, enabling value-sensitive designs that foreground how technology use aligns with personal goals, emotions, and social contexts.
\section{Conclusion}

This paper introduced \ws{}, a lightweight technology probe that scaffolds daily reflection on smartphone use, and presented findings from a two-week deployment with $N$=25 college students. 
Our study uncovered an estimated--actual (E--A) gap in smartphone use, where participants' self-estimations of their device use often diverged from their actual usage patterns.
For instance, we observed statistically significant underestimations in productivity and social app usage and overestimations in entertainment app usage. 
We also examined the relationships between \ea{} and daily self-reported wellbeing. More accurate estimations were associated with higher satisfaction and self-control, but also with greater stress and reduced goal adherence.
Finally, we also found that \ws{} use led to a significant improvement in emotional wellbeing, in terms of positive affect among our participants.
Our interviews revealed that participants recalibrated expectations, recognized recurring patterns, and engaged more intentionally with smartphone use. Their understanding of digital wellbeing expanded beyond time management to include emotional impacts and digital presence, showing that reflection can reframe what wellbeing means in practice. We also observed measurable psychological benefits, including a statistically significant ten percent improvement in positive affect, even within a short two-week window. 
Together, these findings highlight the promise of reflective, user-centered interventions for supporting digital wellbeing. 
By making the often hidden gaps between perception and reality visible, tools like \ws{} can support self-awareness and intentional engagement in everyday digital use. 
Our work contributes to a growing body of research that calls for designing digital wellbeing systems around awareness and meaning, rather than restriction, and points toward new opportunities for creating technologies that empower individuals to develop healthier and more sustainable digital use.




\bibliographystyle{ACM-Reference-Format}
\bibliography{0paper}



\end{document}

\endinput